\documentclass[prb,twocolumn,superscriptaddress,12point,longbibliography]{revtex4-2}
\usepackage{amsmath,amssymb,bm}
\usepackage{wasysym}
\usepackage{graphicx}
\usepackage{epstopdf}
\usepackage{latexsym}
\usepackage{subfigure}
\usepackage[usenames, dvipsnames]{color}
\usepackage[usenames, dvipsnames]{xcolor}
\usepackage{natbib}
\usepackage{braket}
\usepackage{float}
\usepackage[normalem]{ulem}
\usepackage{comment}
\usepackage{mathtools}
\usepackage{array}
\usepackage{multirow}
\usepackage{chemformula}
\usepackage{svg}
\usepackage[T1]{fontenc}
\usepackage[	
citecolor=ForestGreen,
colorlinks, 
linkcolor=violet,
urlcolor=black,
]{hyperref}
\newcommand\redsout{\bgroup\markoverwith{\textcolor{red}{\rule[0.5ex]{2pt}{0.4pt}}}\ULon}
\newcommand\bluesout{\bgroup\markoverwith{\textcolor{black}{\rule[0.5ex]{2pt}{0.4pt}}}\ULon}

\newcommand{\SPhide}[1]{{}}

\begin{document}

\title{\guillemotleft Anticommuting\guillemotright~$\mathbb{Z}_2$ quantum spin liquids}
\author{Sumiran Pujari}
\affiliation{Department of Physics, Indian Institute of
Technology Bombay, Mumbai, MH 400076, India}
\email{sumiran.pujari@iitb.ac.in}
\affiliation{Max Planck Institute for the Physics of Complex Systems, 01187 Dresden, Germany}
\author{Harsh Nigam}
\affiliation{International Centre for Theoretical Sciences, Tata Institute of Fundamental Research, Bangalore 560089, India}

\begin{abstract}
We discuss a class of lattice $S=\frac{1}{2}$ quantum Hamiltonians with bond-dependent Ising couplings and a mutually \guillemotleft anticommuting\guillemotright~algebra of extensively many local $\mathbb{Z}_2$ conserved charges that was explicated in [\href{https://arxiv.org/abs/2407.06236v6}{arXiv:2407.06236}] including the nomenclature. 
This mutual algebra is reminiscent of the spin-$\frac{1}{2}$ Pauli matrix algebra but encoded in the structure of \emph{local conserved charges}.
These models have finite residual entropy density in the ground state with a simple but non-trivial degeneracy counting and concomitant quantum spin liquidity as proved in [\href{https://arxiv.org/abs/2407.06236v6}{arXiv:2407.06236}]. 
The spin liquidity relies on a geometrically site-interlinked character of the local conserved $\mathbb{Z}_2$ charges that is rather natural in presence of an \guillemotleft anticommuting\guillemotright~structure. 
One may contrast this with for example the bond-interlinked character of the local conserved $\mathbb{Z}_2$ charges on the hexagonal plaquettes of the Kitaev honeycomb spin-$\frac{1}{2}$ model which leads to a mutually commuting local algebra. 

In this work, we make several exact statements on the kinds of many-body order that can be present within the class of \guillemotleft anticommuting\guillemotright~quantum spin liquids co-existing with extensive residual ground state entropy. 
We elucidate the differences between the many-body order in these models and that found in some gapped quantum spin liquids whose canonical example is the Kitaev toric code. 
The toric code belongs to the more general class of Levin-Wen or string net constructions that possess mutually commuting algebras for the local conserved charges. 
We also point out a mutually commuting algebra with local support that are naturally expressed as multi-linear Majorana forms in the Kitaev representation of these quantum spin liquids. 
They capture non-trivial quantum resonances throughout the lattice in these \guillemotleft anticommuting\guillemotright~$\mathbb{Z}_2$ quantum spin liquid Hamiltonians.
\end{abstract}
\maketitle

\tableofcontents

\section{Introduction}
\label{sec:intro}

Many-body topological order has been an influential idea in quantum condensed matter physics. 
It provides for example a compelling paradigm for describing phases that are not described by spontaneous symmetry breaking~\cite{Manybody_topo_review_Wen_2013}. 
This has thus played a big role in characterizing (gapped) quantum spin liquid (QSL) phases in the context of quantum magnetism.
In the quantum information processing context, it provides the basis for setting up fault-tolerant quantum computation by putting many physical qubits in a topologically ordered state to form the logical qubits~\cite{Kitaev_2003}. 
The Kitaev toric code~\cite{Kitaev_1997} is the canonical (solvable) example of such ``quantum hardware''.
From a conceptual point of view, it fits into the broader class of quantum many-body (ground) states that are described by the closed loop superposition wavefunction or the string net picture~\cite{Levin_Wen_2005}. 
This is also referred to as Levin-Wen models when the picture is constructive.

Recently, a set of $S=\frac{1}{2}$ models was introduced by one of the authors in Ref.~\cite{GSentropy_preprint,GSentropy_scipost} which has a family resemblance to models with bond-dependent couplings like the Kitaev toric code or the Kitaev honeycomb model~\cite{Kitaev_2006} which also has toric code order as one of the (gapped) many-body topologically ordered phases. 
However the models of Ref.~\cite{GSentropy_scipost} host extensive residual ground state degeneracy with a finite ground state entropy density in stark contrast to known many-body topologically ordered
ground states. 
This is due to an ``\guillemotleft anticommuting\guillemotright''~algebra of local conserved charges that are multi-site spin-$\frac{1}{2}$ operator products which anticommute when they share a site. 
Such an \guillemotleft anticommuting\guillemotright~structure is quite naturally accommodated with bond-dependent couplings. 
The subject of Ref.~\cite{GSentropy_scipost} was to show how this algebraic structure leads to the large residual entropy and quantum spin liquidity without any magnetic or other symmetry broken orders~\cite{Sec2a2b}. 
The \guillemotleft anticommuting\guillemotright~mechanism in fact leads to such large degeneracies for the full spectrum. 
Thus they factor out from the partition function and give a simple additive extensive constant to the Helmholtz free energy independent of temperature. 
It might seem that this situation is not very interesting, but it is actually not so. 
The ground states of these models are not trivial paramagnets but rather entangled states. 
This spectrum level degeneracy is thus simple to count but non-trivial in its consequences.
Ref.~\cite{GSentropy_scipost} exposed how the \guillemotleft anticommuting\guillemotright~mechanism leads to the preceding effects and also compared and contrasted these QSL states to other known QSL states~\cite{Sec2c}.

The goal here is to provide further insights into the structure of these ``\guillemotleft anticommuting\guillemotright~$\mathbb{Z}_2$ quantum spin liquids''.
They will be referred in short to as \guillemotleft ac\guillemotright-$\mathbb{Z}_2$ QSLs in the rest of the paper.
Apart from the precise statements on degeneracy counting and quantum spin liquidity made in the earlier work~\cite{GSentropy_scipost}, we will make the following additional precise statements in this work which also forms an outline of this paper:
\begin{enumerate}
    \item In Sec.~\ref{sec:hidden_charges}, we point out ``hidden'' conserved quantities in the Kitaev Majorana representation with a \emph{commuting} algebra that are present in these \guillemotleft ac\guillemotright-$\mathbb{Z}_2$ QSLs, apart from the \guillemotleft anticommuting\guillemotright~local conserved charges as multi-site spin-$\frac{1}{2}$ operator products mentioned earlier. These quantities are however ``pure gauge'' and \emph{not physical} which has bearing on the lattice gauge theoretic structure of these models as elaborated more in this section.

    \item In Sec.~\ref{sec:emergence_of_many_body_order}, we focus on a square lattice variant with 4-site coupling terms with the Hamiltonian composed directly of the local \guillemotleft anticommuting\guillemotright~conserved charge operators. 
    This is quite in analogy with the Kitaev toric code with its 4-site star and plaquette coupling terms which are also its  (mutually commuting) conserved local charges. 
Kitaev toric code however does not possess the \guillemotleft anticommuting\guillemotright~structure.

We will show that the 4-spin \guillemotleft ac\guillemotright-$\mathbb{Z}_2$ QSL Hamiltonian microscopically maps to a form that is reminiscent of the Wegner's $\mathbb{Z}_2$ Ising gauge theory~\cite{Wegner_1971,Kogut_review_1979} in terms of appropriately defined block spin variables. 
However these variables -- arrived at after a one-time block decimation step described in Sec.~\ref{subsec:block_decimation} -- live on the sites or vertices of the square lattice, and not on the (midpoints of the) bonds or links of the square lattice in a $\mathbb{Z}_2$ Ising gauge theory. 
An equivalent Hamiltonian was in fact studied by Xu and Moore~\cite{Xu_Moore_PRL_2004}.
Thus the 4-spin \guillemotleft ac\guillemotright-$\mathbb{Z}_2$ QSL inherits the unconventional many-body order of the Xu-Moore model in terms of its effective block spin variables. This many-body order co-exists with the residual entropy and the quite featureless spin liquidity of the underlying spin-$\frac{1}{2}$ degrees of freedom.

    \item In Sec.~\ref{sec:superselection}, we give algebraic arguments for the existence of a four-fold degeneracy in the 4-spin \guillemotleft ac\guillemotright-$\mathbb{Z}_2$ QSL. 
    This is analogous to the four-fold topological degeneracy of the Kitaev toric code, but now occurring in absence of a local $\mathbb{Z}_2$ gauge structure for the 4-spin \guillemotleft ac\guillemotright-$\mathbb{Z}_2$ QSL. 
    In Sec.~\ref{sec:star_terms}, we point out the existence of additional quasiparticles or (zero) modes in the 4-spin \guillemotleft ac\guillemotright-$\mathbb{Z}_2$ QSL that are essentially like the ``star'' charges of the Kitaev toric code.
    Additional subtleties related to the topological (or not) nature of the four-fold degeneracy in the 4-spin \guillemotleft ac\guillemotright-$\mathbb{Z}_2$ QSL are discussed in Sec.~\ref{sec:2spin_conservation}.

    \item In Sec.~\ref{sec:non_bipartite}, we point out that all of the above is not specific to bipartite lattices as long as there is a corner-sharing property of the underlying lattice. The corner-sharing property makes accommodating the     local \guillemotleft anticommuting\guillemotright~algebra rather natural for bond-dependent spin-$\frac{1}{2}$ Hamiltonians with ``particle-hole'' symmetric spectra ($\{\pm \epsilon_1, \pm \epsilon_2,\ldots\}$).
    We first do this with the example of a 3-spin \guillemotleft ac\guillemotright-$\mathbb{Z}_2$ QSL
    on the Kagome lattice with corner-sharing triangles. We then discuss a 4-spin \guillemotleft ac\guillemotright-$\mathbb{Z}_2$ QSL pyrochlore lattice with corner-sharing tetrahedra the possibility of which was already hinted at in the conclusion of Ref.~\cite{GSentropy_scipost}.

    \item For the multi-spin \guillemotleft ac\guillemotright-$\mathbb{Z}_2$ QSLs as mentioned in the previously enumerated points on Secs.~\ref{sec:emergence_of_many_body_order},~\ref{sec:superselection} and ~\ref{sec:non_bipartite}, one can make a good amount of progress somewhat like the relation    of the Kitaev toric code to Wegner's $\mathbb{Z}_2$-Ising gauge theory.  
    In Sec.~\ref{sec:2spin_cases}, we discuss the related implications of the above kind of arguments for 2-spin \guillemotleft ac\guillemotright-$\mathbb{Z}_2$ QSLs.

    \item There are additional discussions and a few conjectures in the conclusion Sec.~\ref{sec:conclu}. Verification of these conjectures is left to the future.
\end{enumerate}

\section{\guillemotleft Anticommuting\guillemotright~$\mathbb{Z}_2$ quantum spin liquids in the Kitaev representation}
\label{sec:hidden_charges}

\begin{figure*}
    \centering
    \includegraphics[width=0.8\linewidth]{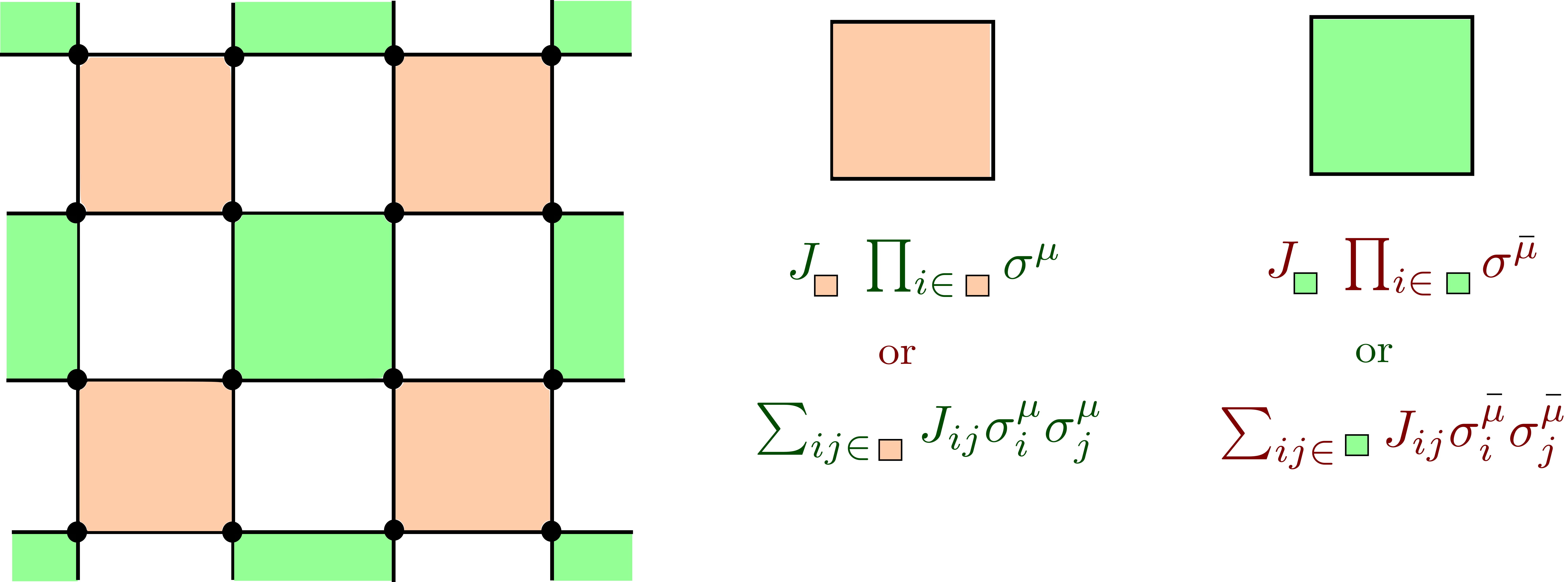}
    \caption{A prototyptical lattice structure for two-dimensional 
    \guillemotleft anticommuting\guillemotright~quantum spin liquid models. 
    Note the ``plaquette-interlinked'' nature of the bond-dependent couplings in this model, when compared to the ``bond-interlinked'' nature of the bond-dependent couplings in the Kitaev honeycomb model along orthogonal directions in spin space.
    Other models say with 3-spin couplings which will have the \guillemotleft anticommuting\guillemotright~algebra can also be easily imagined.
    }
    \label{fig:2dmodel_prototypical}
\end{figure*}

The bond-dependent Hamiltonian model introduced 
in Ref.~\cite{GSentropy_scipost},
\begin{equation}
    H =\; J_x \sum_{\boxed{x}} \left( \sum_{\langle i,j \rangle
    \in\; \boxed{x}} \sigma^x_i \sigma^x_j \right)
    +
    J_z \sum_{\boxed{z}} \left( \sum_{\langle i,j \rangle
    \in\; \boxed{z}} \sigma^z_i \sigma^z_j \right),
    \label{eq:2spin_square_hamil}
\end{equation}
is sketched in Fig.~\ref{fig:2dmodel_prototypical}. 
Note that Eq.~\ref{eq:2spin_square_hamil} assumes translational symmetry and a nearest-neighbour coupling form for the plaquette terms. 
This is however not necessary and there is a much greater variety of local terms in the Hamiltonian that is available while admitting an \guillemotleft anticommuting\guillemotright~structure.
This has been indicated by sketching some other possibilities on the right hand side of Fig.~\ref{fig:2dmodel_prototypical}. 
Eq.~\ref{eq:2spin_square_hamil} can be taken to be the prototypical model with an \guillemotleft anticommuting\guillemotright~structure.
It has finite ground state entropy density and quantum spin liquidity as discussed in Sec.~2.1-2 of Ref.~\cite{GSentropy_scipost}. 
This model does not admit a free fermion solution using the Kitaev representation~\cite{Kitaev_2006}. 
However there are many local conserved charges analogous to the Kitaev honeycomb model. 
They are
\begin{eqnarray}
    \sigma^z_i \sigma^z_j \sigma^z_k \sigma^z_l \text{ on the $\boxed{x}$ plaquettes,}
    \label{eq:4spin_conserved_charge1} \\
    \sigma^x_i \sigma^x_j \sigma^x_k \sigma^x_l \text{ on the $\boxed{z}$ plaquettes.}
    \label{eq:4spin_conserved_charge2}
\end{eqnarray}
The conserved nature of these quantities
may be verified easily. They are essentially conserved (local) $\mathbb{Z}_2$ parities
of these Ising plaquettes. All the above quantities form extensively 
large sets due to their local nature. Unlike the Kitaev toric code or 
Kitaev honeycomb model, these local charges do not all mutually commute. 
The two sets rather \guillemotleft anticommute\guillemotright~with 
each other in the sense described in Sec.~\ref{sec:intro}, i.e., ``multi-site spin-$\frac{1}{2}$ operator products which anticommute when they share a site''. 
Note the single-site sharing property is not strictly necessary as long as there are anticommuting local charges, however the single-site sharing property is quite natural for corner sharing lattices.
This obstructs solvability through the Kitaev approach due to the non-commutation of the local conserved $\mathbb{Z}_2$ charges unlike the Kitaev honeycomb model. 
This \guillemotleft anticommuting\guillemotright structure 
underlies the absence of magnetic or spin ordering~\cite{Sec2a2b}.

If we were to use Kitaev representation~\cite{Kitaev_2006,Fu_Knolle_Perkins_PRB_2018}
($\sigma^\mu_i = i \gamma^0_i \gamma^\mu_j$ with $\gamma$
as Majorana fermions) for Eq.~\ref{eq:2spin_square_hamil}, we would arrive at
\begin{align}
    \label{eq:2spin_square_hamil_kitaev}
    \tilde{H} =\; \sum_{\mu \in \{x,z\}} J_\mu \sum_{\boxed{\mu}} \left( \sum_{\langle i,j \rangle
    \in\; \boxed{\mu}} \left(\gamma^0_i \gamma^0_j \right) 
    \left(\gamma^\mu_i \gamma^\mu_j \right) \right).
\end{align}
The tilde symbol on the left hand side of Eq.~\ref{eq:2spin_square_hamil_kitaev} is to indicate that this operator acts on the extended Hilbert space of four Majorana degrees of freedom per site as per Kitaev's prescription. 
Eq.~\ref{eq:2spin_square_hamil_kitaev} does not reduce to a system of free fermions unlike the Kitaev honeycomb model. 
For the Kitaev honeycomb model, we recall that the $\mathbb{Z}_2$-valued quantities $u^\mu_{ij} \equiv \gamma^\mu_i \gamma^\mu_j$ are remarkably static background gauge fields. 
However for Eq.~\ref{eq:2spin_square_hamil_kitaev}, they do not remain static anymore. 
This is because these $\mathbb{Z}_2$ link or bond quantities coming from within the same plaquette do not all mutually commute anymore, i.e., $[u^\mu_{ij},u^\mu_{jk}]\neq0$. 
This leads to non-commutation with the Hamiltonian as well, thus making them dynamically fluctuating. 
We note here how the \guillemotleft anticommuting\guillemotright~charges of Eq.~\ref{eq:4spin_conserved_charge1},~\ref{eq:4spin_conserved_charge2} appear in this representation
\begin{eqnarray}
    \left( \gamma^0_i \gamma^0_j \gamma^0_k \gamma^0_l
    \right) \left( \gamma^z_i \gamma^z_j \gamma^z_k \gamma^z_l
    \right) \text{ on the $\boxed{x}$ plaquettes,} 
    \label{eq:8majorana_conserved_charge1} \\
    \left( \gamma^0_i \gamma^0_j \gamma^0_k \gamma^0_l
    \right) \left( \gamma^x_i \gamma^x_j \gamma^x_k \gamma^x_l
    \right) \text{ on the $\boxed{z}$ plaquettes,}
    \label{eq:8majorana_conserved_charge2}
\end{eqnarray}
and one can check their non-commuting property
using this representation as well.
Interestingly there are the following additional conserved quantities
\begin{eqnarray}
    \left( \gamma^x_i \gamma^x_j \gamma^x_k \gamma^x_l
    \right) \text{ on the $\boxed{x}$ plaquettes,}
    \label{eq:4majorana_conserved_charge1} \\
    \left( \gamma^z_i \gamma^z_j \gamma^z_k \gamma^z_l
    \right) \text{ on the $\boxed{z}$ plaquettes,}
    \label{eq:4majorana_conserved_charge2}
\end{eqnarray}
with eigenvalues $\pm 1$. 
Furthermore they \emph{mutually commute} with each other and are thus static under Hamiltonian evolution by Eq.~\ref{eq:2spin_square_hamil_kitaev}. 
But these $\mathbb{Z}_2$ quantities do not help to reduce the model to a quadratic fermionic form. 
For the above Majorana multilinears, we take a particular orientation of $i,j,k,l$ on each plaquette, say clockwise starting from a reference site. 
We will adhere to this convention throughout for Majorana multilinears. 

Let us also consider the following 4-spin variant of Eq.~\ref{eq:2spin_square_hamil} composed of the \guillemotleft anticommuting\guillemotright~$\mathbb{Z}_2$ terms themselves. The model is thus
\begin{equation}
    H =\; J_x \sum_{\boxed{x}} \left( \prod_{i \in\; \boxed{x}} 
     \sigma^x_i \right)
    +
    J_z \sum_{\boxed{z}} \left( \prod_{j \in\; \boxed{z}} 
    \sigma^z_j \right),
    \label{eq:4spin_square_hamil}
\end{equation}
again with the following conserved local $\mathbb{Z}_2$ parities
\begin{eqnarray}
    \sigma^z_i \sigma^z_j \sigma^z_k \sigma^z_l \text{ on the $\boxed{x}$ plaquettes,}
    \label{eq:4spin_conserved_charge3} \\
    \sigma^x_i \sigma^x_j \sigma^x_k \sigma^x_l \text{ on the $\boxed{z}$ plaquettes.}
    \label{eq:4spin_conserved_charge4}
\end{eqnarray}
This model and related multi-spin variants will be of special focus in this work. They adhere in a sense to the spirit of Kitaev toric code for which the mutually commuting  terms form the Hamiltonian.
In the Kitaev representation, Eq.~\ref{eq:4spin_square_hamil} becomes
\begin{align}
    \tilde{H} =\;  \sum_{\mu \in \{x,z\}} J_\mu \sum_{\boxed{\mu}}  \left( \prod_{i \in\; \boxed{\mu}} 
     \gamma^0_i \right) \left( \prod_{i \in\; \boxed{\mu}} 
     \gamma^\mu_i \right).
\end{align}
This model also has the additional
mutually commuting Majorana multilinears
\begin{eqnarray}
    \left( \gamma^x_i \gamma^x_j \gamma^x_k \gamma^x_l
    \right) \text{ on the $\boxed{x}$ plaquettes,}
    \label{eq:4majorana_conserved_charge3} \\
    \left( \gamma^z_i \gamma^z_j \gamma^z_k \gamma^z_l
    \right) \text{ on the $\boxed{z}$ plaquettes,}
    \label{eq:4majorana_conserved_charge4}
\end{eqnarray}
as the earlier model in Eq.~\ref{eq:2spin_square_hamil_kitaev}.
Thus we arrive at 
\begin{equation}
    \Tilde{H} =\;  J_x \sum_{\boxed{x}}  \boxed{u_x} 
    \left( \prod_{i \in\; \boxed{x}}  \gamma^0_i \right) 
    \; + \; J_z \sum_{\boxed{z}} \boxed{u_z} 
    \left( \prod_{i \in\; \boxed{x}}  \gamma^0_i \right), 
    \label{eq:4spin_square_hamil_kitaev}
\end{equation}
where $\boxed{u_x}$ and $\boxed{u_z}$ are the conserved $\mathbb{Z}_2$-valued charges in the extended Hilbert space of the Kitaev representation. 
The above is not a quadratic form but rather an interacting Majorana Hamiltonian. 
It is not clear if there is a straightforward exact solvable form for its groundstate and/or eigenspectrum.
We can also check all of the above for 
Eqs.~\ref{eq:2spin_square_hamil},~\ref{eq:4spin_square_hamil} 
in a different representation known as the $SO(3)$ Majorana 
representation~\cite{Tsvelik_1992,Fu_Knolle_Perkins_PRB_2018}
with essentially the same conclusions as above,
i.e., not leading to a free fermion representation
 as is perhaps to be expected.
See Sec. II.A of Ref.~\cite{Fu_Knolle_Perkins_PRB_2018} for
the details on this representation. 

\textbf{Interpretation:} From the above we see that the 4-Majorana conserved quantities written above are $\mathbb{Z}_2$ variables on the plaquettes of the lattice. 
They are thus \emph{unlike} the bond or link $\mathbb{Z}_2$-gauge variables on the bonds of the (say square or honeycomb) lattice as is usual in standard lattice gauge theories. 
They are also gauge dependent variables since we can change their sign using the standard gauge freedom of the Kitaev (or any other) Majorana representation for spin-$\frac{1}{2}$, i.e.,
\begin{align}
\gamma^0_i \rightarrow \epsilon_i \gamma^0_i,\;\;
\gamma^\mu_i \rightarrow \epsilon_i \gamma^\mu_i\;
\;\;\;\text{ keeps }\;\;\;\sigma^\mu_i \rightarrow \sigma^\mu_i 
\\
\text{and implies}
\;\;\;\;\;\;
\boxed{u_\mu} \rightarrow \left( \prod_{i \in \boxed{\mu}} \epsilon_i \right) \boxed{u_\mu},    
\label{eq:kitaev_rep_gauge_freedom}
\end{align}
where $\epsilon_i$ is a site-dependent sign ($\pm 1$). 
Interestingly, while the many-body energetics of the Kitaev honeycomb model depends on the configuration of the background static  $\mathbb{Z}_2$-gauge invariant  \emph{fluxes} -- the $W_p$ conserved quantities of Eq.~6 of Ref.~\cite{Kitaev_2006} --, the many-body energetics of the \guillemotleft anticommuting\guillemotright~$\mathbb{Z}_2$ QSLs under consideration here \emph{can not} depend on the configuration of the plaquette ($\mathbb{Z}_2$) 4-Majorana forms in the background precisely because they are gauge dependent and involve a gauge choice. 

In other words, the Majorana multilinears in Eq.~\ref{eq:4majorana_conserved_charge1},~\ref{eq:4majorana_conserved_charge2},~\ref{eq:4majorana_conserved_charge3},~\ref{eq:4majorana_conserved_charge4} in the context of Eq.~\ref{eq:2spin_square_hamil_kitaev},~\ref{eq:4spin_square_hamil_kitaev}  are conserved quantities only in the extended Hilbert space of the Kitaev Majorana representation, and \emph{not} in the physical Hilbert space of spin-$\frac{1}{2}$ qubits. 
Technically, they do not commute with the site-local projection operator ($D_i \propto \gamma^0_i \gamma^x_i \gamma^y_i \gamma^z_i$) that is necessary to take in account the projective nature of the Kitaev representation~\cite{footnote_projection}. 
These Majorana multilinear symmetries in the extended Hilbert space may however be multiplied or combined to obtain valid symmetries in the physical spin-$\frac{1}{2}$ Hilbert space. 
E.g., for the Kitaev honeycomb model, the 2-Majorana forms $\gamma^\mu_i \gamma^\mu_j$ are gauge dependent quantities that may be static in the extended Hilbert space, while their product around a honeycomb plaquette are $\mathbb{Z}_2$-gauge invariant and define physical fluxes ($W_p$) mentioned before. 
For the case of Eq.~\ref{eq:4majorana_conserved_charge1},~\ref{eq:4majorana_conserved_charge2},~\ref{eq:4majorana_conserved_charge3},~\ref{eq:4majorana_conserved_charge4} and Eq.~\ref{eq:2spin_square_hamil_kitaev},~\ref{eq:4spin_square_hamil_kitaev}, there is no local gauge invariant quantities to be had as opposed to the Kitaev honeycomb model plaquette fluxes. 
However there \emph{is} a \emph{global} gauge invariant quantity which is a product of the conserved 4-Majorana multilinears from all the plaquettes, i.e.,
\begin{align}
\prod_{\mu \in \{x,z\}} \left( \prod_{i \in \boxed{\mu}} \gamma^\mu_i \right)
=
\prod_i \left( \gamma^x_i \gamma^z_i \right) & \label{eq:global_y_Z2} \\
\xrightarrow{\text{projection step } (D_i \equiv~i \gamma^0_i \gamma^x_i \gamma^y_i \gamma^z_i = 1)}\;\;
&
\propto \prod_i \left(i \gamma^0_i \gamma^y_i \right)
= \prod_i \sigma^y_i \nonumber
\end{align}
which is manifestly gauge invariant even before the projection step.
The projection step shows that this is a physical global $\mathbb{Z}_2$ symmetry of Eq.~\ref{eq:2spin_square_hamil_kitaev},~\ref{eq:4spin_square_hamil_kitaev} -- a global $180^\circ$ rotation around the $y$-direction in spin space.
It is in this sense that Eq.~\ref{eq:4majorana_conserved_charge1},~\ref{eq:4majorana_conserved_charge2},~\ref{eq:4majorana_conserved_charge3},~\ref{eq:4majorana_conserved_charge4} capture non-trivial quantum resonances in elementary plaquette motifs as mentioned in the abstract.
We will make a few more remarks on related considerations in the context of the multi-spin \guillemotleft ac\guillemotright-$\mathbb{Z}_2$ QSLs in Secs.~\ref{sec:star_terms} for Eq.~\ref{eq:4spin_square_hamil} and also similar results on non-bipartite lattices in Sec.~\ref{sec:non_bipartite}.

There are several differences when compared to the Kitaev honeycomb model that are worth pointing out:
\begin{enumerate}
    \item We arrived at the plaquette $\mathbb{Z}_2$-gauge dependent flux-unlike variables of Eq.~\ref{eq:4majorana_conserved_charge1},~\ref{eq:4majorana_conserved_charge2},~\ref{eq:4majorana_conserved_charge3},~\ref{eq:4majorana_conserved_charge4} ``directly'' without having had to invoke any underlying ($\mathbb{Z}_2$) gauge group link or bond degrees of freedom. 
    So they can not     clearly be considered as standard gauge-invariant ``fluxes'' from a gauge-theoretic perspective. This shows the non-standard character of these \guillemotleft anticommuting\guillemotright-QSL Hamiltonians when trying to view them as a lattice gauge theory with Majorana matter sector and a $\mathbb{Z}_2$ sector. 

    \item The above point answers to an extent one of the gauge-theoretically oriented structural questions posed in Ref.~\cite{GSentropy_scipost, footnote_gauge_structure}. 
    To elaborate, standard lattice gauge theories are defined with the gauge group variables on the links or bonds (with co-dimension one) of the lattice in any dimension. 
    Products of these gauge group variables around elementary plaquettes (with co-dimension two) are defined to be gauge-invariant fluxes. 
    This is not the state of affairs for Eq.~\ref{eq:4spin_square_hamil_kitaev} or Eq.~\ref{eq:2spin_square_hamil_kitaev}. 
    Interestingly, if we were to take a ``one-dimensional limit'' of Eq.~\ref{eq:2spin_square_hamil} or ~\ref{eq:4spin_square_hamil}, that would lead to the one-dimensional Kitaev chain $H = \sum_i J_x \sigma^x_{i,1} \sigma^x_{i,2} + J_z \sigma^z_{i,2} \sigma^z_{i+1,1}$ with a two-site unit cell which has an \guillemotleft anticommuting\guillemotright~structure as well. 
    This does lead a standard lattice gauge theory with $\mathbb{Z}_2$ gauge variables on the links by virtue of the one-dimensional chain geometry. 
    It turns out that gauge degrees of freedom in one dimension are purely artifactual~\cite{footnote_pure_gauge_artifacts}. 
    For \guillemotleft ac\guillemotright-$\mathbb{Z}_2$ QSLs in higher dimensions, the $\mathbb{Z}_2$ gauge variables are getting assigned to the elementary motifs of the higher dimensional lattice that have the same dimension as the lattice itself -- square plaquettes for the square lattice, tetrahedral motifs for the pyrochlore lattice, etc. -- instead of the bonds or links as in standard lattice gauge theories which are the one-dimensional elementary motifs in any dimension. 
    Thus the \guillemotleft ac\guillemotright-$\mathbb{Z}_2$ QSLs can be considered a \emph{specific} higher-dimensional extension of the Kitaev chain that leads to a non-standard lattice gauge theoretic structure. The other higher-dimensional extensions of the Kitaev chain is of course the Kitaev honeycomb model and its three-dimensional variants on trivalent graphs~\cite{Mandal_Surendran_2009} which lead to a standard lattice gauge theoretic structure. 

    \item The ``matter'' sector (involving only the $\gamma^0$ Majoranas 
    for Eq.~\ref{eq:4spin_square_hamil} 
    but not for Eq.~\ref{eq:2spin_square_hamil}) do not reduce to a
    quadratic hopping Majorana form but rather remain of a 4-Majorana interacting
    form. This non-reduction is true for Eq.~\ref{eq:2spin_square_hamil} or ~\ref{eq:2spin_square_hamil_kitaev} as well.

    We can reinterpret the extensive degeneracies of the \guillemotleft ac\guillemotright-$\mathbb{Z}_2$ QSL~\cite{GSentropy_scipost} through the lens of the Kitaev representation using Eq.~\ref{eq:global_y_Z2}. 
    As discussed earlier, the many-body energetics can not depend on the configuration of $\mathbb{Z}_2$ quantities $\left\{\boxed{u_\mu}\right\}$ associated with the elementary plaquettes. 
    I.e., different configurations of $\left\{\boxed{u_\mu}\right\}$ that are \emph{not} related by the gauge freedom of the Kitaev representation (Eq.~\ref{eq:kitaev_rep_gauge_freedom}) will lead to distinct degenerate states. 
    Thus by counting all possible distinct $\left\{\boxed{u_\mu}\right\}$ configurations that are not gauge copies would yield the degeneracy in a given global parity sector indexed by $\prod_i \sigma^y_i$ that may equal $\pm 1$. This procedure will lead to a count that will be extensive. 
    E.g. a local gauge transformation of the Kitaev representation on a single site $k$, will necessary change the sign of $u_{\boxed{x}}$ \emph{and} $u_{\boxed{z}}$ that border the site $k$.
    Thus we could say take a fixed configuration for $\left\{\boxed{u_\mu}\right\}$, say all of them equal to each other, and then flip the values of only $\left\{\boxed{u_x}\right\}$. 
    The resultant set of $\left\{\boxed{u_\mu}\right\}$ configurations would be distinct and not gauge copies of each other. 
    Clearly the cardinality of this set is extensive.
    
\end{enumerate}

From the above discussion, we saw that the physics of the 4-spin \guillemotleft ac\guillemotright-$\mathbb{Z}_2$ QSL in Eq.~\ref{eq:4spin_square_hamil} is quite different from the many-body physics of Kitaev toric code, and thus more generally, from the physics of Levin-Wen string nets with loop superposition wavefunctions. 
Recall that the toric code is one of the perturbative limits of the Kitaev honeycomb model. 
The 2-spin \guillemotleft ac\guillemotright-$\mathbb{Z}_2$ QSL in Eq.~\ref{eq:2spin_square_hamil} and its non-reduction to a quadratic Majorana form as we saw above also distinguishes itself from the Kitaev honeycomb model that does reduce to hopping Majoranas in the background of static $\mathbb{Z}_2$ fluxes. 
The preceding is noteworthy because the respective Hamiltonians are not entirely dissimilar.

We note here that extensive degeneracy can also be had in models with commuting algebras of local conserved charges. E.g. in Kitaev toric code which resembles that of Eq.~\ref{eq:4spin_square_hamil},
we can set the coefficients of an extensive number of star or plaquette terms to zero~\cite{Ankush_Chaubey_private}.
This is of course to be thought of as zero modes of the well-defined $e$ and $m$ excitations of the toric code. In the \guillemotleft ac\guillemotright-$\mathbb{Z}_2$ QSLs, the status of the existence of well-defined quasiparticles is not quite clear. This was also seen in the preceding discussion on the interacting 4-Majorana form of these models which seems
not easily ``escapable" \emph{including} when 2-spin terms are present such as in Eq.~\ref{eq:2spin_square_hamil_kitaev} or Eq.~\ref{eq:2spin_square_hamil}. 
Similar conjectures were also discussed in Sec.~2.3 of Ref.~\cite{GSentropy_scipost}. 

\section{A 4-spin \guillemotleft  anticommuting\guillemotright~$\mathbb{Z}_2$ quantum spin liquid on the square lattice}
\label{sec:emergence_of_many_body_order}

In this section, we will study further the 4-spin model of Eq.~\ref{eq:4spin_square_hamil} which as mentioned before has a family resemblance to the Kitaev toric code. 
It will be shown that this 4-spin 
\guillemotleft ac\guillemotright-$\mathbb{Z}_2$ QSL is mappable to something that is reminiscent of but not the same as Wegner's $\mathbb{Z}_2$-Ising lattice gauge theory. 
Let us recall that 1) the Kitaev toric code and Wegner's $\mathbb{Z}_2$-Ising gauge theory belong to the same universality class, and 2) matter-less lattice gauge theories (including $\mathbb{Z}_2$-Ising gauge theory) are generally defined on a ``medial'' lattice where the gauge group degrees of freedom live on (the midpoints of) bonds or links of the lattice. 
I.e., there are two gauge variables per unit cell for $\mathbb{Z}_2$-Ising gauge theory on the square lattice as was the original conception of Wegner. 
In our case the mapping step -- to be described soon in Sec.~\ref{subsec:block_decimation} -- will lead to a Hamiltonian which has a similar appearance as the matter-less $\mathbb{Z}_2$-Ising lattice gauge theory but now with the degrees of freedom living on sites or vertices of the (square) lattice and not on bonds of the (square) lattice, i.e., one per unit cell. 
This alternative model has been studied by Xu and Moore~\cite{Xu_Moore_Nucl_Physb_2005} motivated through a different context concerning $p+ip$ superconducting arrays~\cite{Xu_Moore_PRL_2004,Xu_Fu_PRB_2010}. 
It turns out this difference of the degrees of freedom being one or two per unit cell makes a crucial difference as to the nature of the models, in particular whether there exists local gauge transformation generators or not.
On a side note, the Xu-Moore model can also emerge in the toric code in presence of a transverse field~\cite{Vidal_etal_PRB_2009} but not in the Kitaev honeycomb model~\cite{Vidal_Schmidt_Dusuel_PRB_2008}.

For the mapping, we make the following observation: 
the 4-spin plaquette operators $\prod_{i \in \boxed{\mu}}  \sigma^\mu_i$ in terms of their energies do not care about the actual $\{\sigma^\mu_i\}$-spin configuration but only their $\mathbb{Z}_2$-parity in the $\mu$-direction in spin space. 
This $\mathbb{Z}_2$-parity is of course not conserved by the Hamiltonian (Eq.~\ref{eq:4spin_square_hamil}); while the $\mathbb{Z}_2$-parity in $\overline{\mu}$-direction on the $\boxed{\mu}$ plaquette is conserved (Eq.~\ref{eq:4spin_conserved_charge3} and Eq.~\ref{eq:4spin_conserved_charge4}). 
Thus we can effectively treat the non-conserved $\prod_{i \in \boxed{\mu}}  \sigma^\mu_i$ terms on $\left\{\boxed{\mu}\right\}$ plaquette as two-level systems as far as many-body energetics or eigenspectrum is concerned. 

\subsection{One-time block decimation}
\label{subsec:block_decimation}

\begin{figure*}
    \centering
    \includegraphics[width=0.8\linewidth]{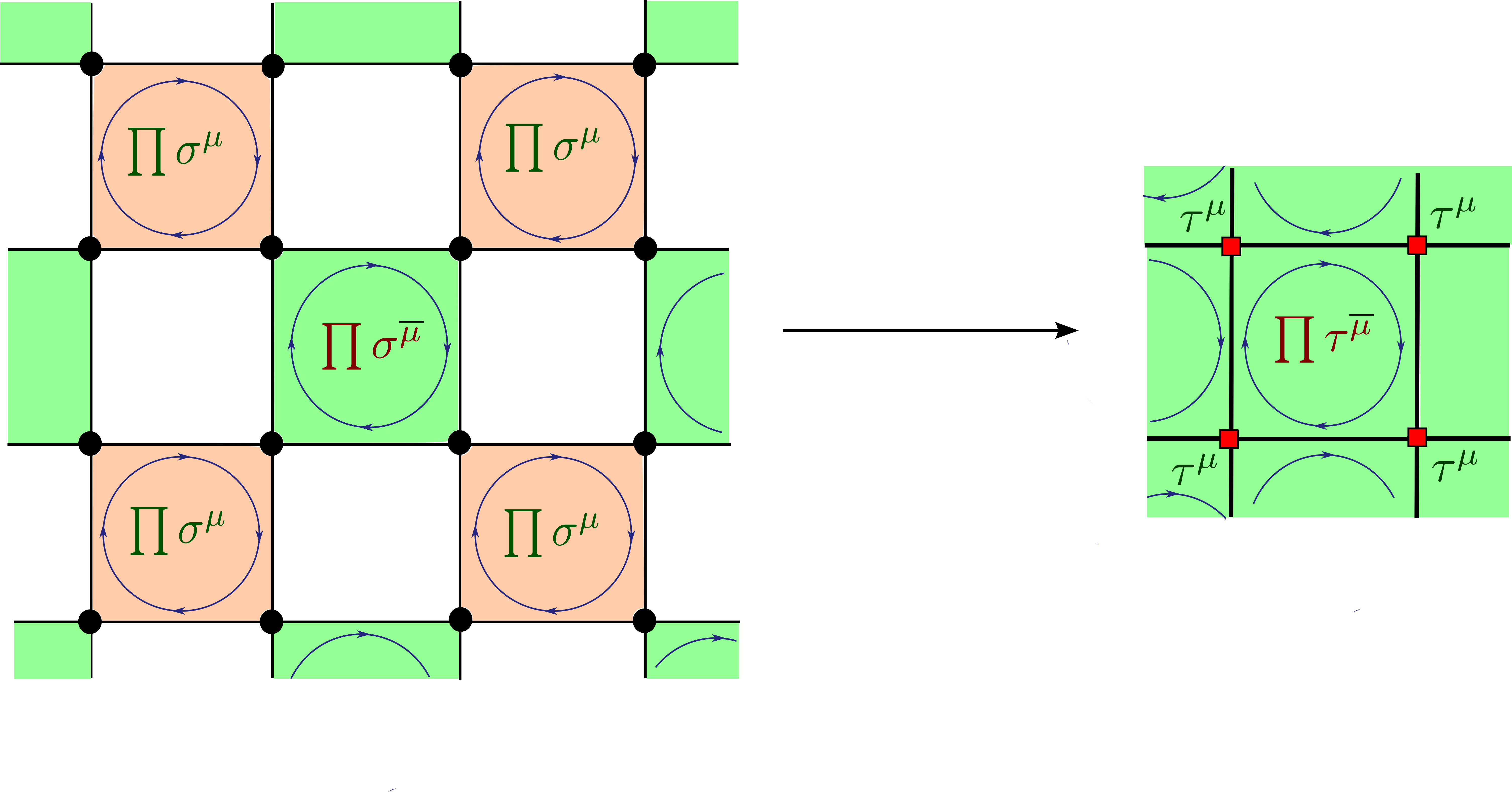}
    \caption{Illustration of the block decimation step
    described in the text of Sec.~\ref{subsec:block_decimation}}
    \label{fig:block_decimation}
\end{figure*}

We can therefore define the following plaquette-parity based two-level variable 
\begin{equation}\label{eq:tau_parity}
    \tau^\mu_I \simeq \prod_{i \in \boxed{\mu}}  \sigma^\mu_i,
\end{equation}
where $I$ indexes the plaquette $\boxed{\mu}$.
It is important to keep in mind that the previous definition
is not an equivalence, but rather is like Kadanoff's block decimation step done ``just once'' to arrive at the plaquette-parity based \emph{two-valued} variable.
I.e., this decimation is a many-to-one map.
To do the decimation step, we choose one of the two sets of 
$\left\{ \boxed{\mu} \right\}$ and $\left\{ \boxed{\overline{\mu}} \right\}$ plaquettes. 
This will lead us to an effecitve square lattice after the decimatiom with the chosen set of plaquettes becoming the sites or vertices of the resultant block decimated lattice
as shown in Fig.~\ref{fig:block_decimation}.

Also as is usual under decimation or coarse-graining, we will lose information in this process -- in this case about the underlying spin configurations of the $\sigma^\mu_i$ variables -- and arrive at an effective model whose Hilbert space has a smaller dimension. It equals $2^{N_\text{uc}}$ after the decimation while we started with $2^{4 \times N_\text{uc}}$ for Fig.~\ref{fig:2dmodel_prototypical} where $N_\text{uc}$ is the number of the unit cells of the original lattice. Under this coarse-graining step, we directly have
\begin{equation}\label{eq:tau_parity_2}
    \sum_{\boxed{\mu}} \left( \prod_{i \in \boxed{\mu}}  \sigma^\mu_i \right)
    \simeq \sum_I \tau^\mu_I
\end{equation}
on the chosen set of plaquettes while organizing the block-decimation or coarse-graining step. 
$I$ denotes the sites or vertices of this one-time block-decimated lattice as shown on the right of Fig.~\ref{fig:block_decimation}. 
For the lattice of Eq.~\ref{eq:4spin_square_hamil} as shown in Fig.~\ref{fig:2dmodel_prototypical}, $I$ is naturally associated to the center of the chosen set of plaquettes $\left\{ \boxed{\mu} \right\}$, and the resultant block-decimated lattice is a square lattice. 
We can actually do this here without \emph{any} loss of information with respect to the many-body eigenspectrum. 
The loss of information happens only for the underlying spin configuration, i.e., the spin configuration info $\{\sigma^\mu_i\}$ for $i \in I$. 
We can afford to do this since we already know that the underlying spin-spin correlations are hyperlocal in space and time and quite featureless as proved in Ref.~\cite{GSentropy_scipost}.

\subsection{Emergence of unconventional many-body order}
\label{subsec:emergence_of_xu_moore}

Now it remains to work out what the other set of plaquette 4-spin terms of the original Hamiltonian map to after the block-decimation.
Since for $\left( \prod_{i \in \boxed{\overline{\mu}}}  \sigma^{\overline{\mu}}_i \right)$, each $\sigma^{\overline{\mu}}_i$ term in the product essentially flips the parity of the $\boxed{\mu}$-plaquette abutting the \boxed{\overline{\mu}}-plaquette at the $i^\text{th}$ site,
therefore we arrive at
\begin{equation}\label{eq:tau_parity_flip}
    \sum_{\boxed{\overline{\mu}}} 
    \left( \prod_{i \in \boxed{\overline{\mu}}}  \sigma^{\overline{\mu}}_i \right)
    \simeq \sum_{\langle IJKL \rangle} \tau^{\overline{\mu}}_I \tau^{\overline{\mu}}_J 
    \tau^{\overline{\mu}}_K \tau^{\overline{\mu}}_L,
\end{equation}
where $\langle IJKL \rangle$ denotes the square plaquettes of the block-decimated
lattice as also indicated in Fig.~\ref{fig:block_decimation}.
Thus the Hamiltonian of Eq.~\ref{eq:4spin_square_hamil} maps under the
above block-decimation to the following:
\begin{equation}
    H \rightarrow H_{bd} = J_\mu \sum_I \tau^\mu_I + 
    J_{\overline{\mu}} \sum_{\langle IJKL \rangle} \tau^{\overline{\mu}}_I \tau^{\overline{\mu}}_J 
    \tau^{\overline{\mu}}_K \tau^{\overline{\mu}}_L,
    \label{eq:block_decimated_hamil}
\end{equation}
which is nothing but the Xu-Moore model. 
The form of Eq.~\ref{eq:block_decimated_hamil} is reminiscent of Wegner's $\mathbb{Z}_2$ Ising gauge theory.
However the decimated block spin or plaquette parity
variables live naturally on the sites or vertices of the square lattice rather on the bonds or links as remarked at the beginning of this section. 
Now we can use the known results for Xu-Moore model to make statements about the 4-spin \guillemotleft ac\guillemotright-$\mathbb{Z}_2$ QSL of Eq.~\ref{eq:4spin_square_hamil} in terms of their plaquette parities which govern the many-body spectrum.
This would be ``on top of'' the $2^{N_{\text{uc}}}$ lower bound on the extensive degeneracy and the quantum spin liquidity stipulated on the underlying the $\sigma^\mu_i$ degrees of freedom by the \guillemotleft anticommuting\guillemotright mechanism.
The following results directly apply to the plaquette parities:
\begin{enumerate}
    \item The Xu-Moore model has an extensive number of symmetries or conserved charges that may be defined on the ``rows'' and ``columns'' of the square lattice. 
    There are thus $2^{L_x + L_y}$ such \emph{non-local} 
    charges. 
    They may be written as $\prod_{I \in {\text{``row''}}} \tau^\mu_I$ and $\prod_{I \in {\text{``column''}}} \tau^\mu_I$ respectively. 
    These have been called as ``sliding'' symmetries in 
    Refs.~\cite{Nussinov_Fradkin_PRB_2005,Xu_Moore_Nucl_Physb_2005}.

    \item These non-local $\mathbb{Z}_2$ symmetries may be spontaneously broken when $\frac{J_{\overline{\mu}}}{J_\mu} > 1$.
    The corresponding order was termed as an ``unconventional bond order'' by Xu and Moore.
    It is captured by non-local order parameters, e.g. through Wilson loops of the form $W(\mathcal{L}) = \prod_{I \in \mathcal{L}} \tau^{\overline{\mu}}_I$.
    For $J_\mu=0$, this order parameter takes $\langle W(\mathcal{L}) \rangle =1$ for any closed loop $\mathcal{L}$.
    For $\frac{J_{\overline{\mu}}}{J_\mu} < 1$, these unconventional orders are absent. 
    See Eq.~18 of Ref.~\cite{Xu_Moore_PRL_2004} and the surrounding discussion. 

    \item The side with unbroken non-local $\mathbb{Z}_2$ symmetries also has a dual form of the same unconventional order. 
    This can be seen through a (self-)duality given on pg.~5 of Ref.~\cite{Xu_Moore_Nucl_Physb_2005}. 
    On the disordered side, the order resides in the variables defined on the dual sites. 
    See Eq.~15 and Eq.~16 of Ref.~\cite{Xu_Moore_Nucl_Physb_2005} for the definitions of the dual variables and the subsequent discussion for the non-local order parameters that capture the many-body order on the disordered side.

    \item The self-duality mentioned above implies a quantum phase transition at $\frac{J_{\overline{\mu}}}{J_\mu} = 1$. This self-duality is 
    also clear at the level of Eq.~\ref{eq:4spin_square_hamil} in terms
    of the chosen set of plaquettes $\left\{\boxed{\mu}\right\}$ or 
    $\{\boxed{\overline{\mu}}\}$ with which to start the block decimation step.
    
    \item It turns out that the Xu-Moore model can also be
    mapped to the $90^\circ$ compass model as shown by Nussinov and Fradkin
    in Ref.~\cite{Nussinov_Fradkin_PRB_2005} which gives another perspective
    on the sliding symmetries.

    \item There are no local gauge transformations for Eq.~\ref{eq:block_decimated_hamil}
    in terms of the $\{\tau^\mu_I\}$ variables where $I,J,K,\ldots$ are the sites of the 
    block-decimated square lattice. 
    We may compare this to what happens in Wegner's $\mathbb{Z}_2$ Ising gauge theory where there
    are rather $2^{L_x L_y}$ number of local conserved charges~\cite{Wegner_1971,Kogut_review_1979,footnote_Wegner}.
\end{enumerate}

\section{Analysis in terms of global superselection sectors}
\label{sec:superselection}

\begin{figure}
    \centering
    \includegraphics[width=0.9\linewidth]{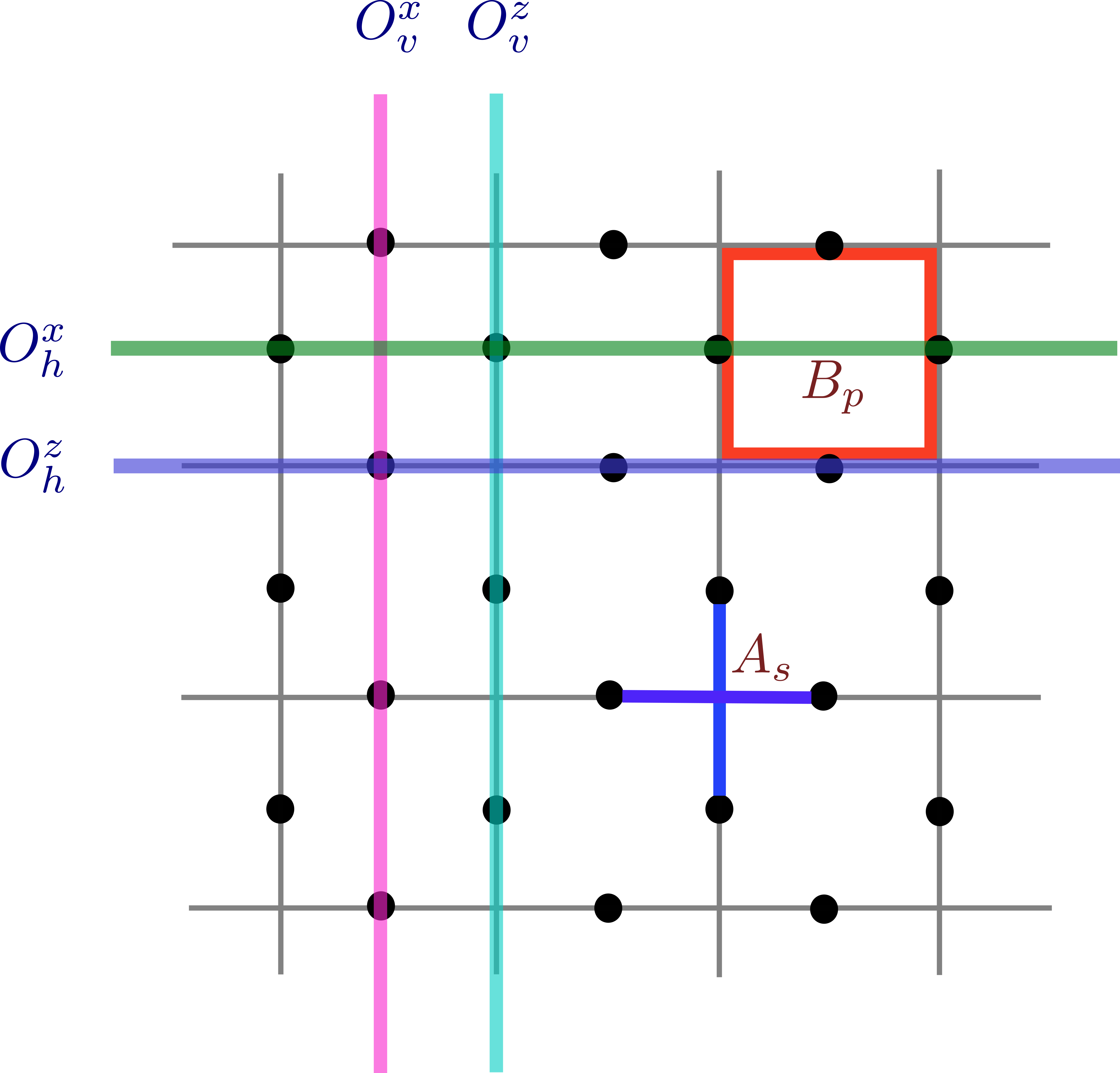}
    \caption{The conserved quantities $O_h^x,~ O_h^z,~ O_v^x,~O_v^z$ for the  Kitaev toric code.}
    \label{fig:ktc_superselection}
\end{figure}

\begin{figure*}
    \centering
    \includegraphics[width=0.6\linewidth]{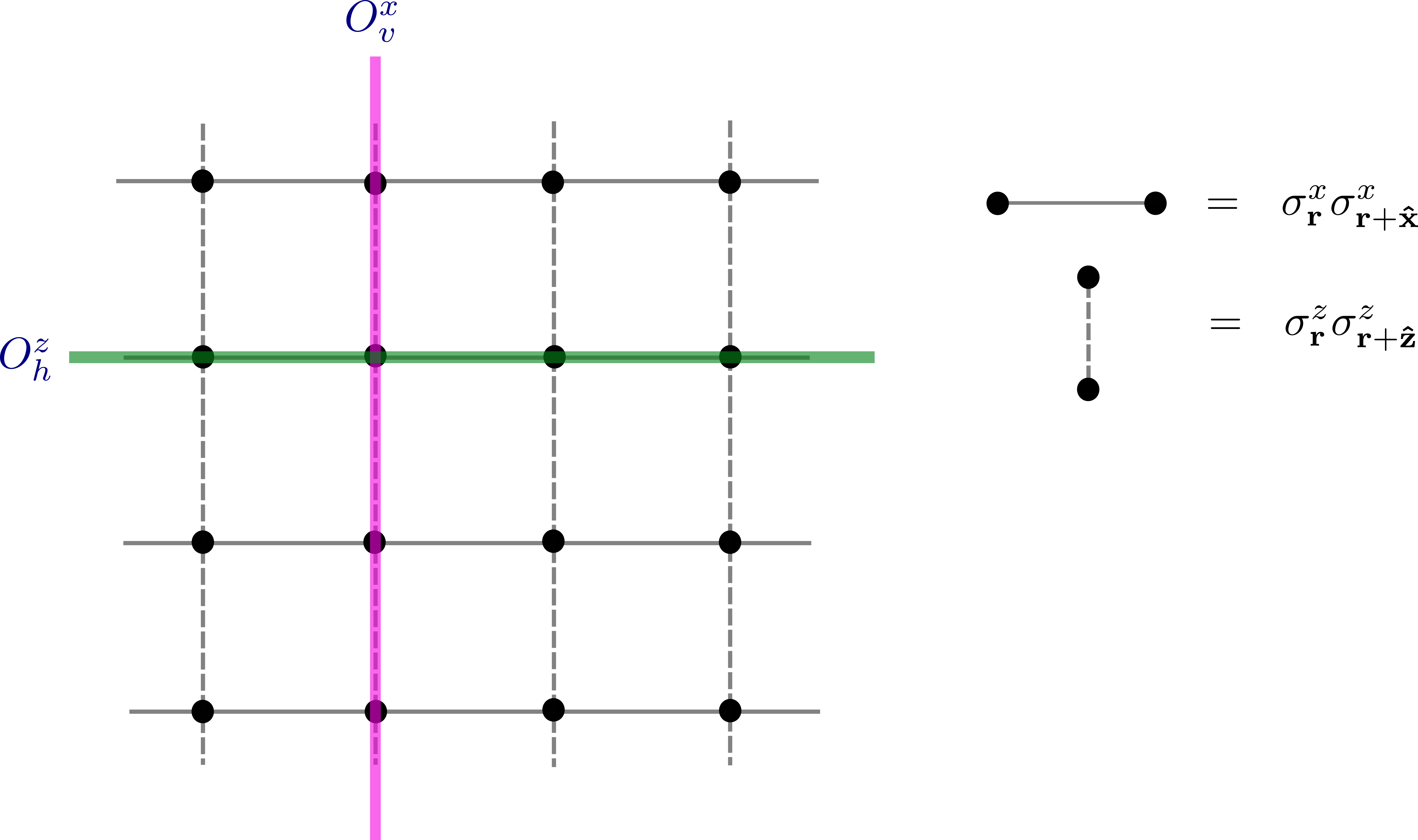}
    \caption{The conserved quantities $O_v^x,~O_h^z$ for the  90$^\circ$ compass model.}
    \label{fig:compass_superselection}
\end{figure*}

In this section, we will investigate the possibility of topologically protected degeracies the 4-spin \guillemotleft ac\guillemotright-$\mathbb{Z}_2$ QSL case given its similarity in appearence to the Kitaev toric code.
Before we do this, we will recall the global ``superselection sector'' arguments for Kitaev toric code (and the 90$^\circ$ compass model) to set the stage for our discussion. 
For the Kitaev toric code model with spin-$\frac{1}{2}$ $\sigma^\mu_i$ operators on the links $\{ i,j,k,\ldots \}$ of the square lattice, the Hamiltonian is
\begin{equation}
    H_{\text{tc}} = -J_s \sum_{s} A^x_s - J_p \sum_{p} B^z_p,  
    \label{eq:kitaev_toric_code}
\end{equation}
where $A^x_s = \prod_{i \in s} \sigma^x_i$ is the so-called star term, $i \in s$ refers to the four links that join at the vertex or star $s$, and $B^z_p = \prod_{i \in p} \sigma^z_i$ is the plaquette term for plaquette $p$.
Apart from the local term $\{A_s\}$ and $\{B_p\}$ that commute with the Hamiltonian, we also have non-local quantities as indicated in Fig.~\ref{fig:ktc_superselection} that commute with the Hamiltonian.
We may convert some of these into each other by appropriately multiplying them with the local conserved operators ($\{A_s\}$ and $\{B_p\}$). 
So considering all of them separately is a form of overcounting of information, and we can restrict ourselves to a minimal set of these non-local conserved operators that are not inter-convertible through multiplication with 
the local conserved operators ($\{A_s\}$ and $\{B_p\}$).
Further details on the inter-convertibility of these non-local operators in App.~\ref{app:interconvertibility}.
One possible choice is $\{O^x_h,O^z_h,O^x_v,O^z_v\}$ as shown in Fig.~\ref{fig:ktc_superselection}:
\begin{eqnarray}
 O_h^\alpha &=& \prod_{i \in \mathcal{L}_h} \sigma_i^\alpha, \label{eq:non-cont-loop-h}\\
 O_v^\alpha &=& \prod_{i \in \mathcal{L}_v} \sigma_i^\alpha, \label{eq:non-cont-loop-v}
\end{eqnarray}
where $\mathcal{L}_h$,~ $\mathcal{L}_v$ are horizontal and vertical non-contractible loops in presence of periodic boundary conditions as shown in Fig.~\ref{fig:ktc_superselection}.
However, they do
not all mutually commute. We have in fact the following:
\begin{eqnarray}
    &[O^x_h,O^z_h]=0, \label{eq:Ohx_Ohz}\\
    &[O^x_v,O^z_v]=0, \label{eq:Ovx_Ovz}\\
    &\{O^x_h,O^z_v\}=0, \label{eq:Ohx_Ovz}\\
    &\{O^x_v,O^z_h\}=0, \label{eq:Ovx_Ohz}\\
    &[O^x_h,O^x_v]=0, \label{eq:Ohx_Ovz_comm}\\
    &[O^z_h,O^z_v]=0. \label{eq:Ohz_Ovz}
\end{eqnarray}
From the above we see that there are four global superselection sectors, say indexed by $\langle O^x_h \rangle = \pm 1$ and $\langle O^z_h \rangle = \pm 1$. 
Because of the existence of the other operators $O^z_v$ and $O^x_v$ not chosen for this indexing that have a mutual \guillemotleft anticommuting\guillemotright~structure, we can again use the degeneracy generation mechanism of \guillemotleft anticommuting\guillemotright~algebras that was exploited at a local level in Ref.~\cite{GSentropy_scipost}. 
Here we will generate a four-fold degeneracy through this mechanism, i.e., for a given eigenstate $|\psi\rangle$ in terms of the above indexing, we get three degenerate orthogonal partner states as tabulated in Tab.~\ref{tab:four_fold_degeneracy}.
\begin{table}[b]
    \centering
    \begin{tabular}{ |m{2cm}||m{1cm}|m{1cm}|m{1cm}|  }
 \hline
  & \;\;$H$ & \;\;$O^x_h$ & \;\;$O^z_h$ \\
 \hline
 \hline
 \;\;$|\psi\rangle$   & \;\;$E$ & \;\;+1 & \;\;+1 \\
 \;\;$|O^z_v\psi\rangle$   & \;\;$E$ & \;\;-1 & \;\;+1 \\
 \;\;$|O^x_v\psi\rangle$   & \;\;$E$ & \;\;+1 & \;\;-1 \\
 \;\;$|O^z_v O^x_v\psi\rangle$   & \;\;$E$ & \;\;-1 & \;\;-1 \\
 \hline
\end{tabular}
    \caption{Four-fold degeneracy generation through the mutually \guillemotleft anticommuting\guillemotright~algebra of the non-local conserved operators in Eq.~\ref{eq:non-cont-loop-h},~\ref{eq:non-cont-loop-v}. The table entries give the eigenvalues of the operators at the top of the table acting on the states listed at the leftmost column of the table.}
    \label{tab:four_fold_degeneracy}
\end{table}
The above non-local \guillemotleft anticommuting\guillemotright~degeneracy is essentially a recapitulation of the proof given in Sec.~3 of Ref.~\cite{Kitaev_2003}. 
In the same section the reader will find the argument for why
this degeneracy may be
considered as topological. This is because these non-local conserved operators 
are not inter-convertible and thus some generic local perturbation -- such as a
transverse field -- can generate splitting between these states 
only at a very high order (proportional to
the system width) in the perturbation.
This implies exponentially small splittings in the system width on general grounds for the above discussed degeneracy.
Such degeneracies with exponentially small splittings in presence of local perturbations \emph{combined} with the absence of spontaneously broken symmetry order described through local order parameters are considered to be hallmarks of topological quantum order. 
For the case of the Kitaev toric code and its associated universality class, this is also called as toric code order.


A similar argument can also be had for the 90$^\circ$ compass model
which actually only leads to a two-fold 
degeneracy~\cite{Doucot_etal_2005,Dorier_Becca_Mila_2005}.
The model is sketched in Fig.~\ref{fig:compass_superselection} and can be written as
\begin{equation}
    H = \sum_{\mathbf{r}} J_x \sigma^x_{\mathbf{r}} \sigma^x_{\mathbf{r}+\mathbf{e_x}} + 
    J_z \sigma^z_{\mathbf{r}} \sigma^z_{\mathbf{r}+\mathbf{e_z}}.
    \label{eq:90deg_compass}
\end{equation}
Recapitulating the technical aspects for completeness, there are no 
local conserved operators for this model,
however there are $2^{2\sqrt{\text{\# unit cells}}}$ non-local conserved operators
corresponding to $\mathbb{Z}_2$-Ising parities for each row and column along appropriate
directions in spin space. 
There is no way to inter-convert between these non-local conserved quantities in this model. 
Yet because of the non-local system-width spanning nature of these conserved $\mathbb{Z}_2$-Ising parities, they are not ``independent'' as far as degeneracy generation mechanism is concerned. 
This would now lead to a different form of overcounting when compared to the Kitaev toric code analysis done above. 
Loosely speaking, the conserved non-local $\mathbb{Z}_2$ Ising parities of the compass
model are all grid-interlocked or cross-linked with each other much more
``strongly'' compared to the conserved local $\mathbb{Z}_2$-Ising parities that are 
plaquette-interlinked to form the \guillemotleft anticommuting\guillemotright~algebra 
in \guillemotleft ac\guillemotright-$\mathbb{Z}_2$ QSLs.
They \emph{are} however independent in terms of the block
diagonalization organization of the compass model Hamiltonian and 
lead to blocks of size 
$\frac{2^{\text{\# unit cells}}}{2^{2\sqrt{\text{\# unit cells}}}}$.
For degeneracy counting, we may thus choose at most only two of the conserved non-local $\mathbb{Z}_2$ parities -- any one of the quantities along 
the two orthogonal direction in real space $O^x_v,O^z_h$ (Fig.~\ref{fig:compass_superselection}). This leads to:
\begin{eqnarray}
    \{O^x_v,O^z_h\}=0. 
\end{eqnarray}
Now, there are only two global superselection sectors, indexed by $\langle O^z_h \rangle = \pm 1$. Since $O^z_h$ anticommutes with $O^x_v$, we can once again invoke the degeneracy-generating mechanism described earlier. This leads to a two-fold degeneracy: for a given eigenstate $|\psi\rangle$, indexed as above, we can construct a degenerate orthogonal partner state as follows: 
\[
\begin{array}{ll}
H |\psi\rangle = E |\psi\rangle , &
O^z_h |\psi\rangle = + |\psi\rangle, \\
H |O^x_v\psi\rangle = E |O^x_v\psi\rangle , &
O^z_h |O^x_v\psi\rangle = - |O^z_v\psi\rangle,
\end{array}
\]
as was shown in Ref.~\cite{Doucot_etal_2005}. 
This two-fold degeneracy will also have exponentially small splittings in presence of additional local perturbations such as a transverse field in the orthogonal $y$-direction. 
However this generally would not be considered a topological degeneracy neither a topologically ordered system, since there is spontaneously broken symmetry order along the $x$- or $z$-direction in spin space (of the Ising type) depending on where $J_x$ or $J_z$ dominates as is intuitive enough.

\begin{figure}
    \centering
    \includegraphics[width=0.8\linewidth]{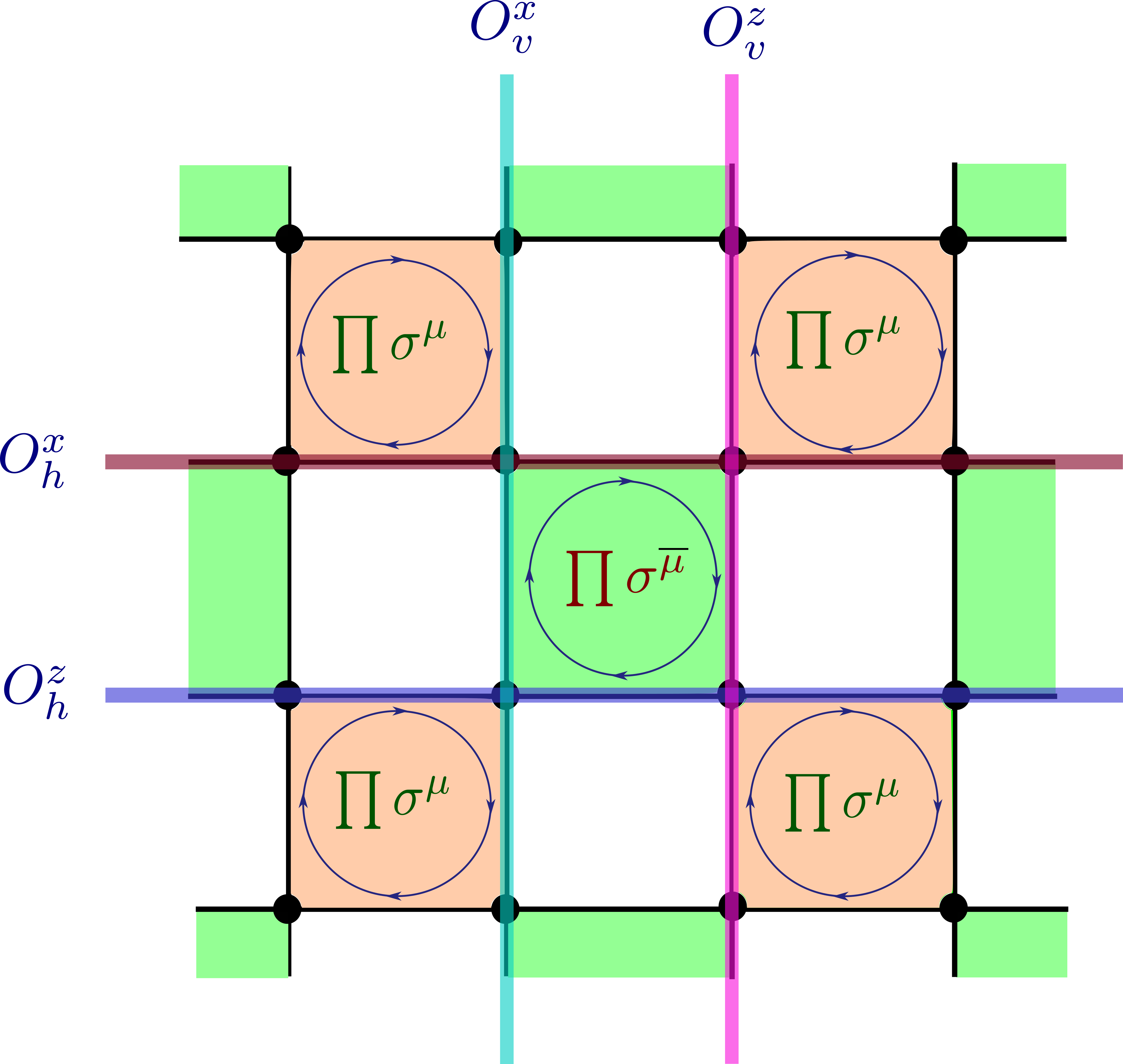}
    \caption{The conserved quantities $O_h^x,~ O_h^z,~ O_v^x,~O_v^z$ for the  4-spin \guillemotleft ac\guillemotright-$\mathbb{Z}_2$ QSL on the square lattice as defined in Eq.~\ref{eq:4spin_square_hamil}.
    We do not consider $O^y_h,~O^y_v$ separately since $O^y_{h/v} = O^z_{h/v} O^x_{h/v}$.
    }
    \label{fig:corner_sq_superselection}
\end{figure}

\subsection{Global superselection and the square lattice
4-spin \guillemotleft ac\guillemotright-$\mathbb{Z}_2$ quantum spin liquid}
\label{subsec:superselection_4spin_square}

We now apply the above line of reasoning to the 4-spin \guillemotleft ac\guillemotright-$\mathbb{Z}_2$ QSL described by Eq.~\ref{eq:4spin_square_hamil}. There are again four non-local conserved quantities analogous to the Kitaev toric code as illustrated in Fig.~\ref{fig:corner_sq_superselection} for this model. Their algebra is given by
\begin{eqnarray}
    &[O^x_h,O^z_h]=0, \\
    &[O^x_v,O^z_v]=0, \\
    &\{O^x_h,O^z_v\}=0, \\
    &\{O^x_v,O^z_h\}=0, \\
    &[O^x_h,O^x_v]=0, \\
    &[O^z_h,O^z_v]=0, 
\end{eqnarray}
which symbolically resembles the algebra of non-local conserved operators in the Kitaev toric code model (cf. Eqs.~\ref{eq:Ohx_Ohz}-\ref{eq:Ohz_Ovz}). 
Even $O^y_h$ and $O^y_v$ are non-local conserved operators, but they do not capture new information since $O^y_{h/v} = O^z_{h/v} O^x_{h/v}$.
Following similar arguments as before, we are again led to a four-fold degeneracy for the 4-spin \guillemotleft ac\guillemotright-$\mathbb{Z}_2$ QSL of Eq.~\ref{eq:4spin_square_hamil}.
This may again be considered topological in nature in the same way as that for 4-fold degeneracy of the Kitaev toric code.
Thus we conclude at the moment that this 4-spin model has unconventional orders in the plaquette parity block spin variables (Sec.~\ref{sec:emergence_of_many_body_order}), quantum spin liquidity in the original spin-$\frac{1}{2}$ variables \emph{and} also a four-fold topological degeneracy formulated in terms of the original spin-$\frac{1}{2}$ $\sigma^\mu_i$ variables as discussed above in this section. 
This is suggestive of a topological order that is distinct than that of the Kitaev toric code that is in the same universality class as the perimeter scaling phase of Wegner's $\mathbb{Z}_2$ Ising gauge theory.
The reason to suspect a distinct topological order is the following: if the four-fold topological degeneracy were to be a consequence of toric code order, one would expect effective star and plaquette terms to emerge in a renormalization group sense after coarse-graining away from the microscopic lattice scale towards larger length scales. 
But as we saw in Sec.~\ref{sec:emergence_of_many_body_order}, the coarse-graining of Eq.~\ref{eq:tau_parity} that was rather natural led to the Xu-Moore model, subsystem symmetries and associated physics.
It is hard to imagine how effective star and plaquette terms would emerge upon further coarse-graining.
We will say some more on this issue at the end of Sec.~\ref{sec:2spin_conservation} that will further highlight why we do not expect the 4-spin square lattice \guillemotleft ac\guillemotright-$\mathbb{Z}_2$ QSL to be in the universality class of the $\mathbb{Z}_2$ Ising gauge theory.

As it stands, there is also another subtlety for the particular 4-spin \guillemotleft ac\guillemotright-$\mathbb{Z}_2$ QSL of Eq.~\ref{eq:4spin_square_hamil} that needs our attention which gives it additional structure \emph{apart from} the \guillemotleft anticommuting\guillemotright~algebra of local conserved $\mathbb{Z}_2$ charges of Eq.~\ref{eq:4spin_conserved_charge3},~\ref{eq:4spin_conserved_charge4}. 
There are additional 2-spin local conserved charges for this model that have definite implications on the global superselection sector structure of Eq.~\ref{eq:4spin_square_hamil}. 
Before we discuss this issue, let us first point out something else analogous to the ``star term'' charges that are well-defined quasiparticle excitations of the Kitaev toric code.

\subsection{``Star term'' charges from ``missing'' plaquettes} 
\label{sec:star_terms}

\begin{figure}
    \centering
    \includegraphics[width=0.7\linewidth]{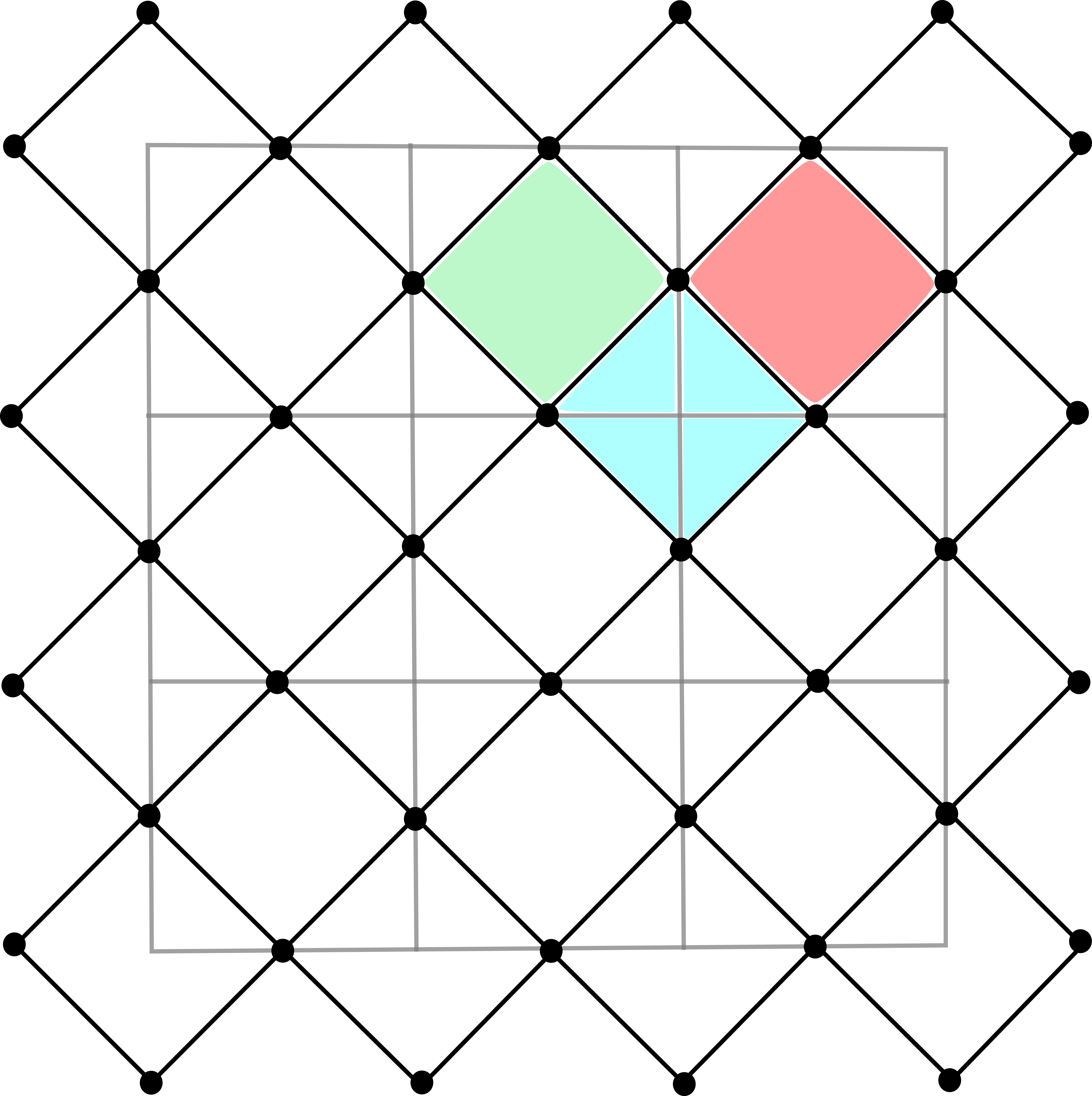}
    \caption{The model in Eq.~\ref{eq:4spin_square_hamil} consists of plaquette terms (red and green) with degrees of freedom located at the sites of a corner-sharing square lattice shown in solid black. One set of the locally conserved plaquette operators are defined on the empty plaquettes (blue), forming the ``star'' terms on the gray square lattice that is $45^\circ$ rotated with respect to the black square lattice. For the gray square lattice, the degrees of freedom reside on the bonds. For details, see Sec.~\ref{sec:star_terms}.}
    \label{fig:square_diamond}
\end{figure}

For the 4-spin \guillemotleft ac\guillemotright-$\mathbb{Z}_2$ QSL of Eq.~\ref{eq:4spin_square_hamil}, apart from the local conserved charges in
Eqns.~\ref{eq:4spin_conserved_charge3},~\ref{eq:4spin_conserved_charge4},~\ref{eq:4majorana_conserved_charge3} and ~\ref{eq:4majorana_conserved_charge4} and the non-local
conserved charges discussed in the previous sections,
there exists another set of local conserved charges. 
These additional local charges are in fact made use of for the inter-convertibility arguments in Sec.~\ref{sec:superselection} and described 
in App.~\ref{app:interconvertibility}. 
They are quantities defined on the ``missing'' plaquettes for Eq.~\ref{eq:4spin_square_hamil}. 
They correspond to the uncoloured plaquettes in Fig.~\ref{fig:2dmodel_prototypical}.
On these plaquettes, we have the following conserved quantities:
\begin{eqnarray}
    \sigma^x_i \sigma^x_j \sigma^x_k \sigma^x_l \text{ on the $\square$ plaquettes,}
    \label{eq:4spin_conserved_charge5} \\
    \sigma^z_i \sigma^z_j \sigma^z_k \sigma^z_l \text{ on the $\square$ plaquettes.}
    \label{eq:4spin_conserved_charge7} 
\end{eqnarray}
Their product, i.e., $\sigma^y_i \sigma^y_j \sigma^y_k \sigma^y_l \text{ on the $\square$ plaquettes}$, is thus also a conserved charge. 
This $\sigma^y_i \sigma^y_j \sigma^y_k \sigma^y_l 
\text{ product on the $\square$ plaquettes}$ can thus be considered as zero-energy quasiparticles or zero modes of the Hamiltonian of 
Eq.~\ref{eq:4spin_square_hamil} since it commutes with Eq.~\ref{eq:4spin_square_hamil}. 
If we considered the original terms of Eq.~\ref{eq:4spin_square_hamil} as (\guillemotleft anticommuting\guillemotright)~plaquette terms analogous to that of the Kitaev toric code, then the additional conserved charges $\sigma^y_i \sigma^y_j \sigma^y_k \sigma^y_l \text{ on the $\square$ plaquettes}$ could be called as the ``star'' terms. 
This is not improper if we were to let the toric code Hamiltonian be defined such that the spins live on the sites or vertices of a square lattice. 
This is slightly different than is the original convention where the spins were assigned to the bonds or links of a square lattice. 
The unit cell defined this way will have halved unit cell area and primitive axis at $45^\circ$ with respect to those in the original convention as illustrated in
Fig.~\ref{fig:square_diamond}.
If we were to add the $\sigma^y_i \sigma^y_j \sigma^y_k \sigma^y_l \text{ product terms on the missing $\square$ plaquettes}$ to the 4-spin \guillemotleft ac\guillemotright-$\mathbb{Z}_2$ QSL Hamiltonian with non-zero coefficients, then these additional charges would also have associated energetics. 

Can we say something about the mutual statistics of the  $\sigma^y_i \sigma^y_j \sigma^y_k \sigma^y_l$ charges on missing plaquettes $\{\square\}$ with respect to the excitations from the other ``occupied'' plaquette sector 
of the 4-spin \guillemotleft ac\guillemotright-$\mathbb{Z}_2$ QSL in 
Eq.~\ref{eq:4spin_square_hamil}? 
Note that the excitations of the 4-spin \guillemotleft ac\guillemotright-$\mathbb{Z}_2$ QSL are described by Xu-Moore physics in terms of the one-time block-decimated plaquette parity variables defined in Sec.~\ref{sec:emergence_of_many_body_order}.
This is after discounting the extensive degeneracy stipulated by the \guillemotleft anticommuting\guillemotright~mechanism.
Another point of note which will become relevant in the following Sec.~\ref{sec:2spin_conservation} is that the extensive degeneracy stipulated by the \guillemotleft anticommuting\guillemotright~mechanism can also be argued through the lens of the additional symmetries on the missing plaquettes of Eq.~\ref{eq:4spin_conserved_charge5},~\ref{eq:4spin_conserved_charge7} pointed out above, alternatively to what was done using those of Eq.~\ref{eq:4spin_conserved_charge3},~\ref{eq:4spin_conserved_charge4} (or equivalently Eq.~\ref{eq:4spin_conserved_charge1},~\ref{eq:4spin_conserved_charge2}) in Ref.~\cite{GSentropy_scipost}.

We furthermore note for completeness that in the Kitaev representation, the following 4-Majorana 
forms from the missing plaquettes $\{\square\}$ are conserved quantities in the extended Hilbert space as well:
\begin{eqnarray}
    \left( \gamma^0_i \gamma^0_j \gamma^0_k \gamma^0_l
    \right) \text{ on the $\square$ plaquettes,}
    \label{eq:4majorana_conserved_charge5} \\
    \left( \gamma^x_i \gamma^x_j \gamma^x_k \gamma^x_l
    \right) \text{ on the $\square$ plaquettes,}
    \label{eq:4majorana_conserved_charge6} \\
    \left( \gamma^y_i \gamma^y_j \gamma^y_k \gamma^y_l
    \right) \text{ on the $\square$ plaquettes,}
    \label{eq:4majorana_conserved_charge7} \\
    \left( \gamma^z_i \gamma^z_j \gamma^z_k \gamma^z_l
    \right) \text{ on the $\square$ plaquettes.}
    \label{eq:4majorana_conserved_charge8}
\end{eqnarray}
These are also not all independent due to the projective
constraint of the Kitaev Majorana representation. See Section 4.1
of Ref.~\cite{Kitaev_2006}.

\subsection{Additional 2-spin local conserved charges}
\label{sec:2spin_conservation}

\SPhide{Addition of star term leads to topological degeneracy, putting $\sigma_y$ field at all sites and doing pertubation theory gives star terms.}

There exists \emph{additional} local conserved quantities for the particular Hamiltonian in Eq.~\ref{eq:4spin_square_hamil} that were not noted down till now intentionally for the purpose of presentation. 
They are composed of 2-spin terms given by:
\begin{eqnarray}
    &\sigma_i^x \sigma_j^x~~~\text{for}~i,j \in \boxed{z}, \label{eq:2spin_symm1}\\
    &\sigma_i^z \sigma_j^z~~~\text{for}~i,j \in \boxed{x}.  \label{eq:2spin_symm2}
\end{eqnarray}
All other local symmetry operators previously defined in Eqs.~\ref{eq:4spin_conserved_charge3},~\ref{eq:4spin_conserved_charge4},~\ref{eq:4spin_conserved_charge5}, and~\ref{eq:4spin_conserved_charge7} can in fact be constructed using combinations of these two-spin operators~\cite{Diptiman_private}. 
However, this is a particularity of the the 4-spin \guillemotleft ac\guillemotright-$\mathbb{Z}_2$ QSL of Eq.~\ref{eq:4spin_square_hamil}. 
For the 2-spin \guillemotleft ac\guillemotright-$\mathbb{Z}_2$ QSL of Eq.~\ref{eq:2spin_square_hamil}, there are no such 2-spin terms as conserved local charges. 
In this sense, the local conserved $\mathbb{Z}_2$-Ising partities of Eqs.~\ref{eq:4spin_conserved_charge3},~\ref{eq:4spin_conserved_charge4} have a certain primacy and is the prime structure with regards to \guillemotleft ac\guillemotright-$\mathbb{Z}_2$ QSLs. 
Later in this section, we will write down a more general Hamiltonian still with 4-spin coupling terms for which the above 2-spin operators of Eq.~\ref{eq:2spin_symm1},~\ref{eq:2spin_symm2} are not going to be remain conserved. 
This will give another perspective on the enlarged symmetry of Eq.~\ref{eq:4spin_square_hamil} through Eqs.~\ref{eq:2spin_symm1},~\ref{eq:2spin_symm2}. 

Given the existence of the 2-spin conserved local $\mathbb{Z}_2$ charges in Eq.~\ref{eq:2spin_symm1},~\ref{eq:2spin_symm2}, one needs to redo the degeneracy counting via the \guillemotleft anticommuting\guillemotright~mechanism. 
The details of doing such degeneracy counting was discussed using the examples of 2-spin \guillemotleft ac\guillemotright-$\mathbb{Z}_2$ QSLs in Sec.~2.1 of Ref.~\cite{GSentropy_scipost}. 
(Strictly speaking the the \guillemotleft anticommuting\guillemotright~mechanism gives a lower bound on the spectral degeneracies). 
If we were to use the \guillemotleft anticommuting\guillemotright~conserved charges of Eqs.~\ref{eq:4spin_conserved_charge3},~\ref{eq:4spin_conserved_charge4}, then we would conclude a degeneracy of $2^{\#\text{unit cells}}$ since there is one pair of \guillemotleft anticommuting\guillemotright~conserved charges per unit cell. 
This is also true for Eq.~\ref{eq:2spin_square_hamil} via Eqs.~\ref{eq:4spin_conserved_charge1},~\ref{eq:4spin_conserved_charge2}. But given the 2-spin conserved operators of Eqs.~\ref{eq:4spin_conserved_charge3},~\ref{eq:4spin_conserved_charge4}, there are now more conserved quantities within the unit cell. 
The counting now leads to a lower bound of $8^{\#\text{unit cells}}$. 

In brief, the argument goes as follows: We choose one of the mutually commuting sets of \guillemotleft anticommuting\guillemotright~conserved charges, say the set of $\sigma_i^x \sigma_j^x$ for $i,j \in \boxed{z}$ (cf. Eq.~\ref{eq:2spin_symm1}). 
Using the  \guillemotleft anticommuting\guillemotright~$\mathbb{Z}_2$ operators from the other set of $\sigma_i^z \sigma_j^z$ for $i,j \in \boxed{x}$ (cf. Eq.~\ref{eq:2spin_symm2}), we can start generating distinct degenerate states. 
The application of any one of the 
\guillemotleft anticommuting\guillemotright~$\sigma_i^z \sigma_j^z$ from a given $\boxed{x}$ leads to a distinct degenerate partner eigenstate.
This is because it leads to an unique flip in the pattern of $\sigma_i^x \sigma_j^x$ values in the affected neighbouring $\boxed{z}$. 
Similarly, simultaneous application of multiple $\sigma_i^z \sigma_j^z$ terms coming from one or more unit cells will lead to other distinct degenerate partner eigenstates with distinct flipping patterns in the $\sigma_i^x \sigma_j^x$ values in the affected $\boxed{z}$s. 
This leads to the lower bound of $8^{\#\text{unit cells}}$ on the degeneracy generated through the \guillemotleft anticommuting\guillemotright~mechanism.

Importantly, we also see that the non-local operators discussed in the previous section (Eqs.~\ref{eq:non-cont-loop-h},~\ref{eq:non-cont-loop-v}) can also be generated by appropriate products of the local 2-spin symmetry operators in Eqs.~\ref{eq:2spin_symm1} and~\ref{eq:2spin_symm2}. 
This decomposition in terms of local symmetries is not possible for the non-local operators of the Kitaev toric code shown in Fig.~\ref{fig:ktc_superselection}. 
This leads to the different topological sectors being disconnected under toric code Hamiltonian evolution~\cite{footnote_XChen_youtube}. 
In this sense, the 4-fold degeneracy of Hamiltonian in Eq.~\ref{eq:4spin_square_hamil} is not topologically protected as thus not the same as that of the toric code and its topological order.
We give the argument for the lack of topological protection due to the aforesaid local decomposition of the non-local conserved operators in an appendix for completeness (App.~\ref{app:TD_details}).

\begin{figure}
    \centering
    \includegraphics[width=0.8\linewidth]{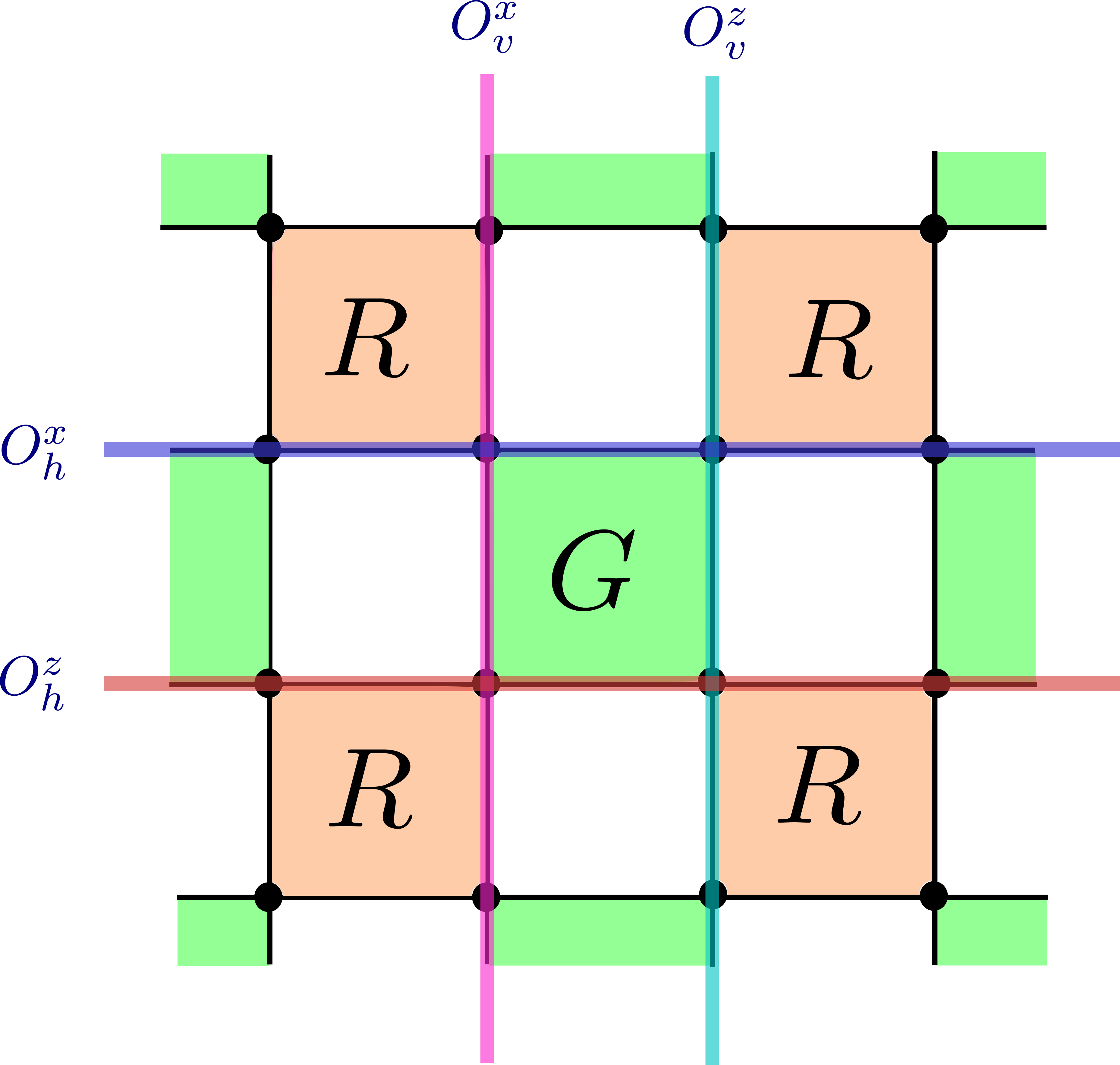}
    \caption{A sketch of the model in Eq.~\ref{eq:Big_4spin_model} defined on a corner-sharing square lattice with conserved quantities $O_h^x,~ O_h^z,~ O_v^x,~O_v^z$. It defines a family of 4-spin \guillemotleft ac\guillemotright-$\mathbb{Z}_2$ QSLs which contains the 4-spin \guillemotleft ac\guillemotright-$\mathbb{Z}_2$ QSL of Eq.~\ref{eq:4spin_square_hamil} as one of its members. }
    \label{fig:BIG_sq_model}
\end{figure}

Thus what do we make of the 4-fold degeneracy of Eq.~\ref{eq:4spin_square_hamil} on the torus that must be 
there via the \guillemotleft anticommuting\guillemotright~non-local operators of Fig.~\ref{fig:corner_sq_superselection} discussed in Sec.~\ref{subsec:superselection_4spin_square}. 
To answer this in a ``corner-sharing'' way, we write down a more general 4-spin model that voids the 2-spin local conserved charges that are inescapable for Eq.~\ref{eq:4spin_square_hamil}. 
It is as follows:
\begin{equation}
    \label{eq:Big_4spin_model}
    H_{\text{gen1}} =  \sum_{\mu \in \{R,G\}} \left[ \sum_{\boxed{\mu}} \left( \alpha_\mu \prod_{i \in\; \boxed{\mu}} 
   \sigma^x_i + \beta_\mu \prod_{i \in\; \boxed{\mu}} 
   \sigma^z_i \right) \right].
\end{equation}
Here, $R$ and $G$ label the corner-sharing plaquettes as depicted in Fig.~\ref{fig:BIG_sq_model} with red and green coloring respectively. 
For the choice $\alpha_R = \beta_G = 0$ or $\alpha_G = \beta_R = 0$, the model reduces to the Hamiltonian in Eq.~\ref{eq:4spin_square_hamil}. 
Also $\alpha_R = \alpha_G = 0$ (or $\beta_R = \beta_G = 0$) reduces to the Kitaev toric code without the star term, i.e., Eq.~\ref{eq:kitaev_toric_code} with $J_s = 0$.
The model in Eq.~\ref{eq:Big_4spin_model} does not possess any 2-spin local conserved quantities of the form given in Eqs.~\ref{eq:2spin_symm1} and~\ref{eq:2spin_symm2}.  
As a consequence, the non-local operators $O_h^x$, $O_h^z$, $O_v^x$, and $O_v^z$ shown in Fig.~\ref{fig:BIG_sq_model} cannot be constructed from local conserved quantities. 
Furthermore they are themselves conserved under or commute with the Hamiltonian of Eq.~\ref{eq:Big_4spin_model}. 
This implies that there is genuine topological degeneracy for Eq.~\ref{eq:Big_4spin_model} with exponentially small level splitting in presence of local perturbations such as a transverse field. 

The symmetries of Eq.~\ref{eq:Big_4spin_model} are defined \emph{only} on the ``missing'' plaquettes (the white plaquettes in Fig.~\ref{fig:BIG_sq_model}). 
They are of the same form as those given in Eqs.~\ref{eq:4spin_conserved_charge5} and~\ref{eq:4spin_conserved_charge7}. 
There are no conserved operators on the $\left\{\boxed{R}\right\}$ and $\left\{\boxed{G}\right\}$ plaquettes except in the limit of $\alpha_R = \beta_G = 0$ or $\alpha_G = \beta_R = 0$. 
However we can use the \guillemotleft anticommuting\guillemotright~algebra present in Eqs.~\ref{eq:4spin_conserved_charge5},~\ref{eq:4spin_conserved_charge7} to again argue for quantum spin liquidity and residual entropies~\cite{GSentropy_preprint}. 
Given a) the lack of spontaneously broken symmetry order of the 4-spin \guillemotleft ac\guillemotright-$\mathbb{Z}_2$ QSL of Eq.~\ref{eq:4spin_square_hamil}, b) the presence of the non-local \guillemotleft anticommuting\guillemotright~conserved charges and thus an associated 4-fold degeneracy on a torus that are however locally decomposable in terms of the 2-spin charges of Eq.~\ref{eq:2spin_symm1},~\ref{eq:2spin_symm2}, and c) that Eq.~\ref{eq:4spin_square_hamil} is the $\alpha_R = \beta_G = 0$ or $\alpha_G = \beta_R = 0$ limit of Eq.~\ref{eq:Big_4spin_model}, we conclude that the particular 4-spin \guillemotleft ac\guillemotright-$\mathbb{Z}_2$ QSL of Eq.~\ref{eq:4spin_square_hamil} is sitting at the boundary of a more general topologically ordered phase. 
This topologically ordered phase is captured representatively through the 4-spin \guillemotleft ac\guillemotright-$\mathbb{Z}_2$ QSLs of Eq.~\ref{eq:Big_4spin_model} where the topological degeneracy will again co-exist with extensive residual entropies and quantum spin liquidity. 
The \guillemotleft anticommuting\guillemotright~structure of Eq.~\ref{eq:Big_4spin_model} as discussed above would stipulate again a lower bound of $2^{\#\text{unit cells}}$ on the extensive degeneracy due to the absence of the 2-spin conserved operators similar to the 2-spin \guillemotleft ac\guillemotright-$\mathbb{Z}_2$ QSL of Eq.~\ref{eq:2spin_square_hamil}.

Of course given the presence of the conserved quantities analogous to the toric code star or $e$ charges discussed in Sec.~\ref{sec:star_terms}, we can also perform the modification in an ``edge-sharing'' way by including the 4-$\sigma^y$ product terms on the missing plaquettes in the Hamiltonian, i.e.,
\begin{equation}\label{eq:Big1_4spin_model}
H_{\text{gen2}} =  
\sum_{\mu \in \{x,z\}} \left[ J^\mu \sum_{\boxed{\mu}} \left( \prod_{i \in\; \boxed{\mu}} \sigma^\mu_i \right) \right]
    +
    J^y \sum_{\square} \left( \prod_{i \in\; \square} \sigma^y_i \right).
\end{equation}
or, more generally, $H_{\text{gen2}}=H_{\text{gen1}} + J^y \sum_{\square} \left( \prod_{i \in\; \square} \sigma^y_i \right)$. Each individual summand in the last term of Eq.~\ref{eq:Big1_4spin_model} are now well-defined quasiparticles  excitations instead of being zero modes as in the discussion of Sec.~\ref{sec:star_terms}. There are several aspects of Eq.~\ref{eq:Big1_4spin_model} worth pointing out:
\begin{itemize}
    \item Firstly, as was the purpose of the modification, the 2-spin conserved quantities of Eq.~\ref{eq:2spin_symm1},~\ref{eq:2spin_symm2} are voided, i.e., not conserved anymore, due to the presence of the last term in Eq.~\ref{eq:Big1_4spin_model}. 
    Eq.~\ref{eq:Big1_4spin_model} still possesses the same non-local algebra as Eq.~\ref{eq:Big_4spin_model} and thus is robustly topologically ordered phase with a four-fold degeneracy on the torus. 
    Eq.~\ref{eq:4spin_square_hamil} can again be thought of as living on the boundary of the topological phase described by Eq.~\ref{eq:Big1_4spin_model}.

    \item Secondly, $G_\square \equiv  \prod_{i \in\; \square} \sigma^y_i$ is in fact a local symmetry ($G_\square H G^{-1}_\square = H \;\forall\;\text{missing plaquettes }\{\square\}$) and can be considered as the gauge charge generator of the star or $e$ charges analogous to the toric code. Given this, in the spirit of the relation between the toric code and the $\mathbb{Z}_2$ Ising gauge theory, we may now write down the following Hamiltonian
    \begin{equation}
       H_{\text{gen3}} =  \sum_{\mu \in \{x,z\}} J^\mu \sum_{\boxed{\mu}} \left( \prod_{i \in\; \boxed{\mu}} \sigma^\mu_i \right)
    +
    J^y \sum_{i} \sigma^y_i 
    \label{eq:ac_z2_ising_gauge_theory}
    \end{equation}
    which also possesses the same local symmetry coming from $G_\square$. 
    This can be considered an \guillemotleft anticommuting\guillemotright~variant of Wegner's $\mathbb{Z}_2$ Ising gauge theory. 
    I.e., its charge sector generated by $G_\square$ is exactly the same as that of the $\mathbb{Z}_2$ Ising gauge theory. 
    However, the magnetic or flux sector of the $\mathbb{Z}_2$ Ising gauge theory is not present for the above Hamiltonian.
    In fact the conjugate sector to the charge sector is now the 4-spin \guillemotleft ac\guillemotright~$\mathbb{Z}_2$ QSL whose excitations are nothing like that of the magnetic or flux charges of the toric code or the $\mathbb{Z}_2$ Ising gauge theory.
    Recall that the many-body energetics of the 4-spin \guillemotleft ac\guillemotright~$\mathbb{Z}_2$ QSL was captured by an effective Xu-Moore model in terms of the plaquette parity variables of Eq.~\ref{eq:tau_parity} as worked out in Sec.~\ref{subsec:emergence_of_xu_moore}. 
    The last term of Eq.~\ref{eq:Big1_4spin_model} (i.e. $G_\square$) will not generate any non-trivial operator action on the plaquette parities, i.e., $G_\square$ will be proportional to the identity operator in terms of the $\{\tau_I\}$ variables following a similar logic as that of Eq.~\ref{eq:tau_parity_flip}.
    As had been posed earlier in Sec.~\ref{sec:star_terms}, the mutual statistics of the charges from the two sectors is thus also not clear.
    One expects that Eq.~\ref{eq:Big1_4spin_model} will be perturbatively generated from Eq.~\ref{eq:ac_z2_ising_gauge_theory} in the limit of $\frac{J^y}{J^\mu} \ll 1$.

    \item Finally for Eq.~\ref{eq:Big1_4spin_model} which is an \guillemotleft anticommuting\guillemotright~version of the toric code, we can ask if there is a pictorial representation similar to the loop superposition wavefunctions for the toric code~\cite{footnote_XChen_youtube} and more generally string nets~\cite{Levin_Wen_2005,Lin_Levin_Burnell_2021}. 
    Because of the local symmetry afforded by $\{G_\square\}$, we can work in the $\{\sigma^y\}$-basis and try to set up a loop superposition picture since these gauge charges are well-defined, i.e., setting $G_\square=1\;\forall\;\square$ to remain in the ground state manifold of the gauge charge sector will automatically lead to loops in the $\{\sigma^y\}$-basis similar to string nets. 
    However because of the \guillemotleft anticommuting\guillemotright~nature of the conjugate sector, the loop superposition will necessarily be different than string nets. 
    For the solvable cases of string nets, the loop superposition becomes especially simple, i.e., an uniform superposition over all closed loop configurations~\cite{footnote_XChen_youtube}.
    This will not be the case in presence of \guillemotleft anticommuting\guillemotright~algebras.
    Whether there is still an identifiable structure is an open, interesting question.
    Also from the above perspective, Eq.~\ref{eq:4spin_square_hamil} being the $J^y=0$ limit of Eq.~\ref{eq:Big1_4spin_model} will contain open strings as well in the superposition, which is alternatively stated as the gauge charges becoming zero modes of the theory (cf. Sec.~\ref{sec:star_terms}).
\end{itemize}

We end this section with the observation that the Majorana representation of the 4-spin \guillemotleft ac\guillemotright-$\mathbb{Z}_2$ QSL Hamiltonian (Eq.~\ref{eq:4spin_square_hamil_kitaev}) also admits analogous 2-Majorana forms as symmetry operators:
\begin{eqnarray}
    &\gamma_i^x \gamma_j^x~~~\text{for}~i,j \in \boxed{x}, \label{eq:2spin_symm3}\\
    &\gamma_i^z \gamma_j^z~~~\text{for}~i,j \in \boxed{z}.  \label{eq:2spin_symm4}
\end{eqnarray}
These 2-Majorana conserved operators are also mutually commuting like the 4-Majorana symmetry operators defined in Eqs.~\ref{eq:4majorana_conserved_charge3},~\ref{eq:4majorana_conserved_charge4},~\ref{eq:4majorana_conserved_charge6}, and thus do not have an \guillemotleft anticommuting\guillemotright~algebra. 
The 4-Majorana symmetry operators can straightforwardly be constructed or recovered from the local two-Majorana symmetry operators in Eqs.~\ref{eq:2spin_symm3} and~\ref{eq:2spin_symm4}.


\section{\guillemotleft Anticommuting\guillemotright~$\mathbb{Z}_2$ quantum spin liquids on non-bipartite lattices}
\label{sec:non_bipartite}

From the discussions so far, we see that the essential algebraic structure underpinning the \guillemotleft ac\guillemotright-$\mathbb{Z}_2$ QSLs does not fundamentally rely on the bipartite nature of the supporting lattice. 
Instead, it is the corner-sharing geometry of the supporting lattice that plays a critical role in enabling the construction of bond-dependent spin-$\frac{1}{2}$ Hamiltonians with non-commuting local conserved quantities. 
Lattices with elementary motifs that share a common site thus are natural supporting lattices. 
The square plaquettes of the square lattice formed the focus till now as in the previous sections. 
The non-bipartite generalizations to be discussed in this section further demonstrate the flexibility of the \guillemotleft anticommuting\guillemotright~framework and its potential relevance to a broader class of frustrated quantum magnets, e.g. the pyrochlore construction can be termed as a ``$\mathbb{Z}_2$ quantum spin ice'' by stretching the ice terminology somewhat.

In this section, we illustrate how the \guillemotleft ac\guillemotright-$\mathbb{Z}_2$ QSL framework or model construction naturally extends to non-bipartite lattices possessing corner-sharing motifs. 
In particular, we will consider two main examples: 1) The kagome lattice with corner-sharing triangles where the \guillemotleft anticommuting\guillemotright~algebra of conserved charges will be \emph{absent} in the elementary triangular motifs. 
But it will be \emph{present} in the elementary hexagonal motifs thereby making the results of Ref.~\cite{GSentropy_scipost} applicable including quantum spin liquidity in the lattice spin variables. 
2) The pyrochlore lattice with corner-sharing tetrahedra which admits an \guillemotleft anticomutting\guillemotright~algebraic structure directly on the elementary tetrahedral motifs and the concomitant physics quite like the square lattice models discussed in the previous sections. 
In the following sections, we will focus on the greater-than-2-spin coupling constructions similar to the bipartite square case in the previous two sections. 
For the 2-spin coupling constructions, the results of Sec.~\ref{sec:hidden_charges} will of course still apply; we will also discuss the 2-spin non-bipartite variants some more in the subsequent Sec.~\ref{sec:2spin_cases}.

\subsection{Kagome lattice}
\label{subsec:kagome}

\begin{figure*}
    \centering
    \includegraphics[width=0.75\linewidth]{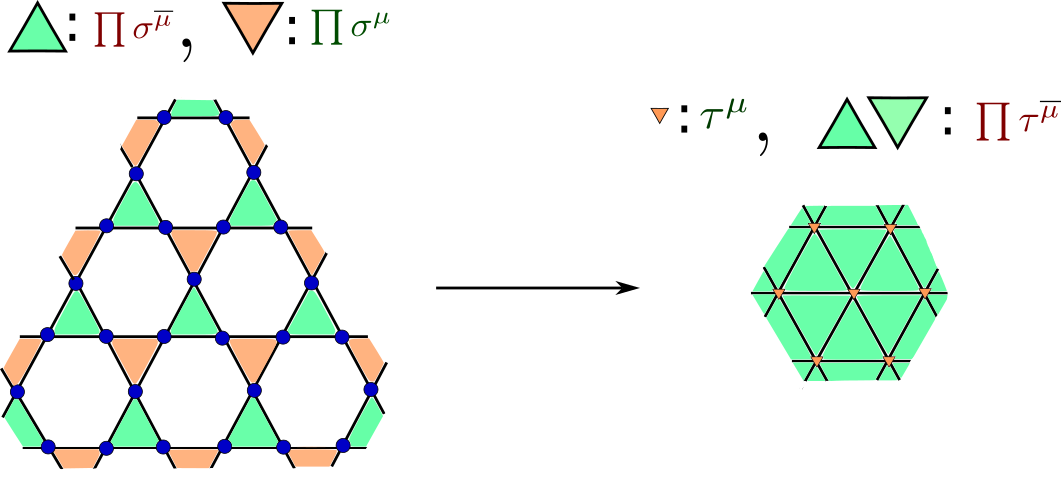}
    \caption{Illustration of the block decimation step for the kagome lattice as set up in Sec.~\ref{subsubsec:block_decimation_kagome}}
    \label{fig:block_decimation_Kagome}
\end{figure*}

We write down a 3-spin coupling model on the Kagome lattice, analogous to Eq.~\ref{eq:4spin_square_hamil}, as shown in the left panel of Fig.~\ref{fig:block_decimation_Kagome}. 
The Kagome lattice having the corner-sharing property naturally accommodates such a model. 
It can be written as
\begin{equation}
    H =\; J_x \sum_{\bigtriangledown} \left( \prod_{i \in\; \bigtriangledown} 
     \sigma^x_i \right)
    +
    J_z \sum_{\triangle} \left( \prod_{j \in\; \triangle} 
    \sigma^z_j \right).
    \label{eq:3Kagome_Ham}
\end{equation}
Note the above Hamiltonian is not invariant under the time reversal operation on the spin-$\frac{1}{2}$ moments ($\mathcal{T} \sigma^\mu_i \mathcal{T}^{-1} = - \sigma^\mu_i$), unlike the square lattice based models discussed till now and the pyrochore model to be discussed later. In Kitaev's Majorana representation, Eq.~\ref{eq:3Kagome_Ham} becomes
\begin{equation}
    \Tilde{H} =\;  J_x \sum_{\bigtriangledown}  \left( \prod_{i \in\; \bigtriangledown} 
     i \gamma^0_i \gamma^x_i \right) 
    +
     J_z \sum_{\triangle} \left( \prod_{i \in\; \triangle} 
     i \gamma^0_i \gamma^z_i \right).
    \label{eq:3Kagome_Ham2}
\end{equation}    
As in Sec.~\ref{sec:hidden_charges}, the tilde symbol on the left hand side of Eq.~\ref{eq:3Kagome_Ham2} is again to indicate that this operator acts on the extended Hilbert space of 4 Majorana degrees of freedom per site as per Kitaev's prescription. 
Here the 3-Majorana form analogs of the conserved 4-Majorana forms in Eq.~\ref{eq:4majorana_conserved_charge1} and related equations now defined on the occupied triangular plaquettes ($\triangle$, $\bigtriangledown$) are not conserved quantities,
\begin{equation}\label{eq:3Kagome_majorana_commut1}
    \left[\prod_{i \in \bigtriangledown}\gamma_i^x,~ \tilde{H}  \right] \neq 0,~~~~~ \left[\prod_{i \in \triangle}\gamma_i^z,~ \tilde{H}  \right] \neq 0.
\end{equation}
Such $n$-Majorana form symmetries are in fact absent for plaquettes composed of an odd number of bonds. However, Majorana multilinear symmetries  defined on the ``missing'' hexagonal plaquettes of the Kagome lattice are indeed present. The corresponding symmetry operators are
\begin{eqnarray}
    \prod_{i \in \hexagon}\gamma^x_i,~~~~~ \prod_{i \in \hexagon}\gamma^y_i, ~~~~~ \prod_{i \in \hexagon}\gamma^z_i,~~~~~ \prod_{i \in \hexagon}\gamma^0_i, \label{eq_gamma_xz_hex} 
\end{eqnarray}
where $\hexagon$ denotes the hexagonal plaquette of the kagome lattice.  All of these operators are symmetries in the extended Hilbert space of Kitaev Majorana fermions by definition, 
\begin{align}\label{eq:3Kagome_majorana_commut2}
    \left[\prod_{i \in \hexagon}\gamma_i^x,~ \tilde{H}  \right] = 0,~~~~~ \left[\prod_{i \in \hexagon}\gamma_i^y,~ \tilde{H} \right] = 0, \nonumber \\ \left[\prod_{i \in \hexagon}\gamma_i^z,~ \tilde{H}  \right] = 0,~~~~~ \left[\prod_{i \in \hexagon}\gamma_i^0,~ \tilde{H}  \right] = 0
\end{align}
but \emph{do not} commute with the projection operator $D_i \propto i \gamma^0_i \gamma^x_i \gamma^y_i \gamma^z_i$ that projects us back into the physical Hilbert space.  Thus they are \emph{not} physical symmetries or conserved charges.
\SPhide{@Harsh, do these 6-Majorana forms commute with the projection-to-the-physical-space operator ($D_i \propto \gamma^0_i \gamma^x_i \gamma^y_i \gamma^z_i$)? HN: NO}
However, their product yields $\prod_{i \in \hexagon}\sigma^x_i$, $\prod_{i \in \hexagon}\sigma^y_i$, and $\prod_{i \in \hexagon}\sigma^z_i$ which are symmetry operators in the physical Hilbert space, i.e.,
\begin{equation}\label{eq:Kagome_sigmay_commut1}
    \left[\prod_{i \in \hexagon}\sigma_i^x,~ H  \right] = 0,~~~~~\left[\prod_{i \in \hexagon}\sigma_i^y,~ H  \right] = 0,~~~~~\left[\prod_{i \in \hexagon}\sigma_i^z,~ H  \right] = 0.
\end{equation}
But they do not all mutually commute and rather form an \guillemotleft anticommuting\guillemotright~algebra supported on the hexagonal plaquettes! 
Thus spin liquidity and a residual ground state entropy is guaranteed~\cite{GSentropy_scipost}. 
We may still give a well-chosen subset of these operators some energetics to obtain quasiparticles in an exact fashion on the hexagonal motifs similar to the star term discussion in Sec.~\ref{sec:star_terms}. 
Note that unlike the case of (corner-sharing) square lattice (cf. Eq.~\ref{eq:2spin_square_hamil},\ref{eq:4spin_square_hamil} and their respective conserved plaquette operators Eq.~\ref{eq:4spin_conserved_charge1},~\ref{eq:4spin_conserved_charge2},~\ref{eq:4spin_conserved_charge3},~\ref{eq:4spin_conserved_charge4}), in the case of Kagome lattice here which is a corner-sharing triangular lattice, Eq.~\ref{eq:3Kagome_Ham} does not admit \guillemotleft anticommuting\guillemotright~local conserved operators on the elementary triangular plaquettes. 
This is a direct consequence of the fact that these ``occupied'' plaquettes of the Kagome lattice are composed of an odd number of bonds. In particular, the following operators do not commute with the Hamiltonian:
\begin{equation}\label{eq:Kagome_sigma_symm}
    \left[\prod_{i \in \triangle} \sigma_i^x, H \right] \neq 0,~~~~~ \left[\prod_{i \in \bigtriangledown} \sigma_i^z, H \right] \neq 0,
\end{equation}
neither do the corresponding 3-Majorana forms as mentioned in Eq.~\ref{eq:3Kagome_majorana_commut1}. 

\begin{figure}[b]
    \centering
    \includegraphics[width=0.8\linewidth]{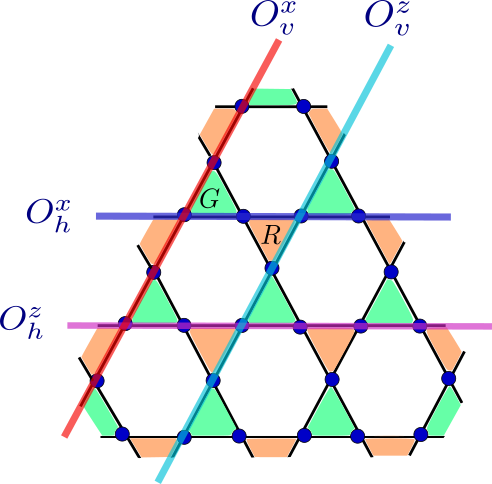}
    \caption{The conserved quantities $O_h^x,~ O_h^z,~ O_v^x,~O_v^z$ for the model Hamiltonian defined in Eq.~\ref{eq:3Kagome_Ham} and Eq.~\ref{eq:Big_Kag_3spin_model} on the kagome lattice.}
    \label{fig:Kagome_superselection}
\end{figure}

\begin{figure*}
    \centering
    \includegraphics[width=0.6\linewidth]{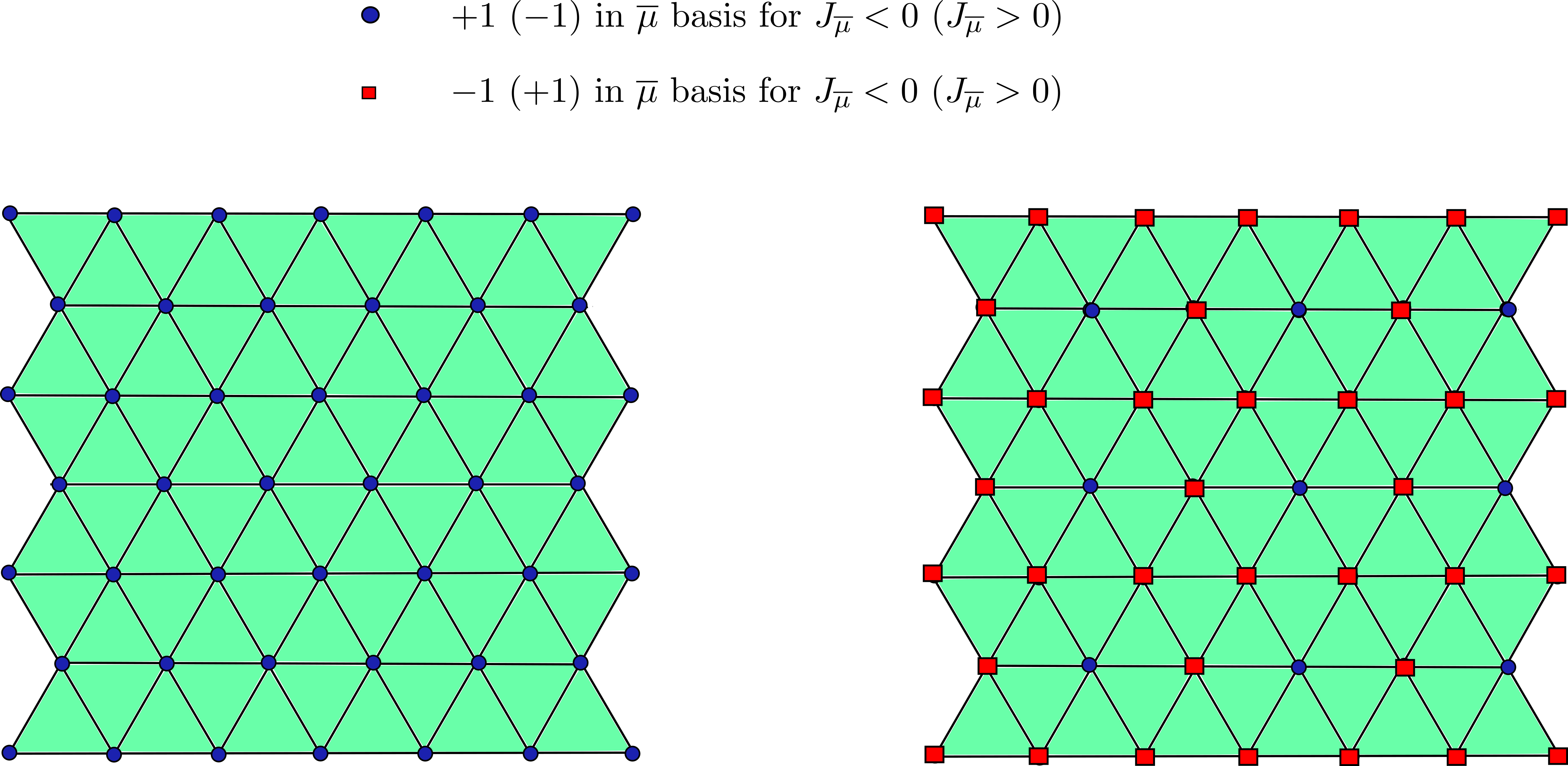}
    \caption{Schematic of the ferromagnetic and one of the three-sublattice long-range ordered states for Xu-Moore model on triangular lattice when $J_\mu=0$.}
    \label{fig:Kagome_LRO}
\end{figure*}

We note that Eq.~\ref{eq:Kagome_sigmay_commut1} and ~\ref{eq:Kagome_sigma_symm} are not true for a 2-spin coupling version of Eq.~\ref{eq:3Kagome_Ham}, i.e.,
\begin{equation}
    H =\; J_x \sum_{\bigtriangledown} \left( \sum_{\langle i,j \rangle \in\; \bigtriangledown} 
     \sigma^x_i \sigma^x_j \right)
    +
    J_z \sum_{\triangle} \left( \sum_{\langle i,j \rangle \in\; \triangle} 
    \sigma^z_i \sigma^z_j \right).
    \label{eq:3Kagome_Ham}
\end{equation}
In fact, the equality counterpart of the inequality in Eq.~\ref{eq:Kagome_sigma_symm} is true for Eq.~\ref{eq:3Kagome_Ham} above leading to an \guillemotleft anticommuting\guillemotright~algebraic structure on the elementary triangular plaquettes, and thus would share properties with the 2-spin \guillemotleft ac\guillemotright-$\mathbb{Z}_2$ QSL on the square lattice (Eq.~\ref{eq:2spin_square_hamil}) discussed in Sec.~\ref{sec:hidden_charges}.

For the 3-spin case of Eq.~\ref{eq:3Kagome_Ham}, there continues to exist two-spin and two-Majorana local symmetry operators of the form discussed in Eqs.~\ref{eq:2spin_symm1}–\ref{eq:2spin_symm4}, within the triangular plaquettes as follows:
\begin{eqnarray}
    &\sigma_i^x \sigma_j^x~~~\text{for}~i,j \in \triangle, \label{eq:2spin_Kag_symm1}\\
    &\sigma_i^z \sigma_j^z~~~\text{for}~i,j \in \bigtriangledown,  \label{eq:2spin_Kag_symm2}
\end{eqnarray}
and in the Majorana representation (Eq.~\ref{eq:3Kagome_Ham2}):
\begin{eqnarray}
    &\gamma_i^x \gamma_j^x~~~\text{for}~i,j \in \bigtriangledown, \label{eq:2spin_Kag_symm3}\\
    &\gamma_i^z \gamma_j^z~~~\text{for}~i,j \in \triangle,  \label{eq:2spin_Kag_symm4}
\end{eqnarray}
with implications analogous to those described in Sec.~\ref{sec:2spin_conservation}.
The non-local quantities $ O_h^x,~ O_h^z,~ O_v^x,~ O_v^z $, defined along non-contractible loops on the torus (Eqs.~\ref{eq:non-cont-loop-h},~\ref{eq:non-cont-loop-v} and Fig.~\ref{fig:Kagome_superselection}) are also present as conserved non-local charges, and are again decomposable in terms of the local 2-spin symmetries in Eq.~\ref{eq:2spin_Kag_symm1},~\ref{eq:2spin_Kag_symm2}. Consequently, this model does not exhibit genuine topological order in the toric code sense. However, and again mirroring the discussion in Sec.~\ref{sec:2spin_conservation}, we can define a modified Hamiltonian on the Kagome lattice as
\begin{equation}\label{eq:Big_Kag_3spin_model}
    H_{\text{gen1}} =  
    \sum_{\mu \in \{R,G\}}
    \sum_{\triangle_\mu} \left( \alpha_\mu \prod_{i \in\; \triangle_\mu} 
   \sigma^x_i + \beta_\mu \prod_{i \in\; \triangle_\mu} 
   \sigma^z_i \right)
\end{equation}
where each triangle (labelled $ R $ and $ G $ in Fig.~\ref{fig:Kagome_superselection}) hosts both types of terms,
or
\begin{equation}\label{eq:Big1_Kag_3spin_model}
    H_{\text{gen2}} =  
    \sum_{\mu \in \{x,z\}}
    \sum_{\triangle_\mu} \left( J^\mu \prod_{i \in\; \triangle_\mu} 
   \sigma^\mu_i \right)
   +
   \sum_{\hexagon} \left( J^y \prod_{i \in\; \hexagon} 
   \sigma^y_i \right)
\end{equation}
or $H_\text{gen1}+\sum_{\hexagon} \left( J^y \prod_{i \in\; \hexagon} 
   \sigma^y_i \right)$ most generally.
These generalized Kagome models lacks any two-spin local conserved quantities, and the non-local operators $ O_h^x,~ O_h^z,~ O_v^x,~ O_v^z $ now obey the same algebra as described in Eqs.~\ref{eq:Ohx_Ohz}–\ref{eq:Ohz_Ovz}. 
Furthermore, they also have an \guillemotleft anticommuting\guillemotright~algebra on the hexagonal plaquettes as well given by Eq.~\ref{eq:Kagome_sigmay_commut1}.
Thus using arguments similar to those in the previous sections, we conclude that these 3-spin Kagome models realizes an \guillemotleft ac\guillemotright-$ \mathbb{Z}_2 $ quantum spin liquid with four-fold topological ground state degeneracy. 
Finally, the 3-spin model of Eq.~\ref{eq:3Kagome_Ham} being 
appropriate limits of Eq.~\ref{eq:Big_Kag_3spin_model} or Eq.~\ref{eq:Big1_Kag_3spin_model} sits at the boundary of a genuinely topologically ordered phase with its topological order being distinct from toric code order in line with Sec.~\ref{subsec:superselection_4spin_square} and ~\ref{sec:2spin_conservation}.
We also write down for completeness the following Hamiltonian
\begin{equation}
    H_{\text{gen3}} =  
    \sum_{\mu \in \{x,z\}}
    \sum_{\triangle_\mu} \left( J^\mu \prod_{i \in\; \triangle_\mu} 
   \sigma^\mu_i \right)
   +
   J^y \sum_i \sigma^y_i 
\end{equation}
which is expected to generate Eq.~\ref{eq:Big1_Kag_3spin_model} perturbatively when $\frac{J^y}{J^\mu} \ll 1$ following the discussion of Sec.~\ref{sec:2spin_conservation}.

\begin{figure*}
    \centering
    \includegraphics[width=0.8\linewidth]{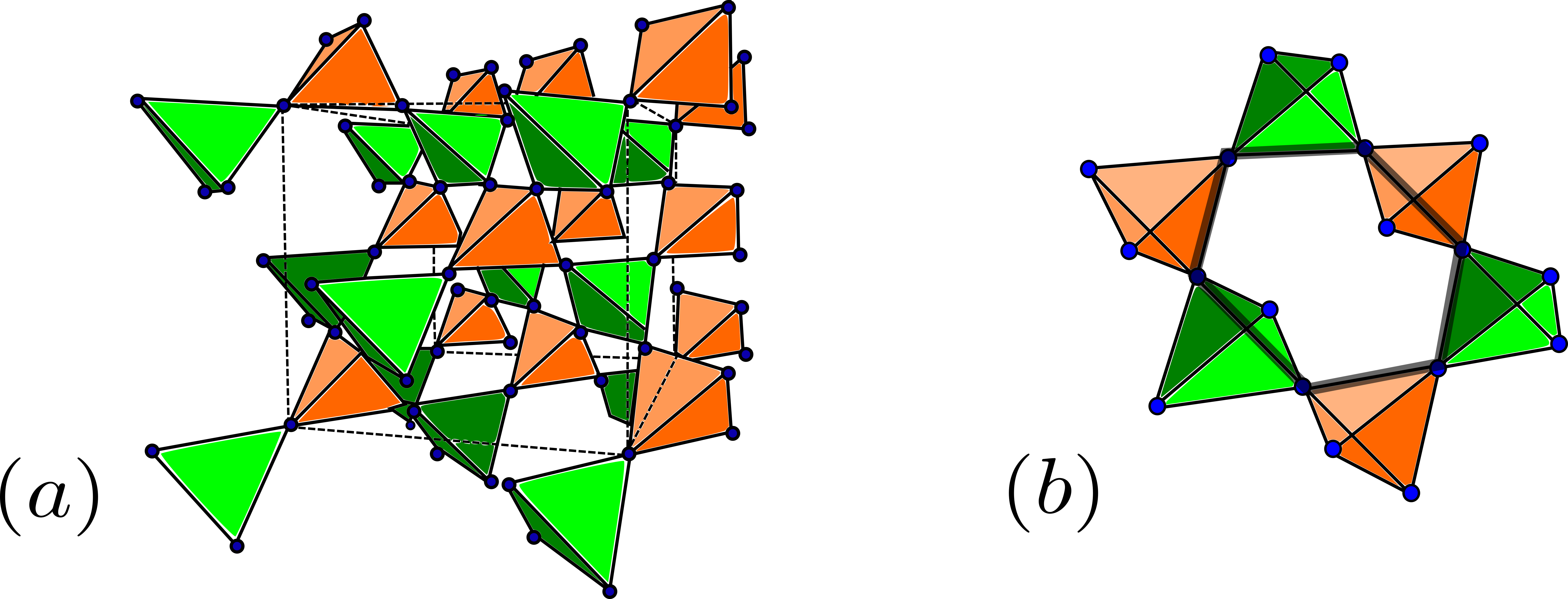}
    \caption{Panel (a) shows a schematic of a pyrochlore lattice built upon a face-centred cubic (FCC) lattice. Panel (b) shows a hexagonal plaquette enclosed by six contiguous tetrahedra of the pyrochlore lattice.
    }
    \label{fig:pyro_loop}
\end{figure*}

\subsubsection{Symmetry breaking order in the 3-spin model}
\label{subsubsec:block_decimation_kagome}

Similar to the block decimation done for the 4-spin model on the square lattice in Sec.~\ref{subsec:block_decimation}, we can perform a  coarse-graining procedure on the Kagome lattice Hamiltonian given in Eq.~\ref{eq:3Kagome_Ham}. 
As before we can define a plaquette-parity variable analogous to Eq.~\ref{eq:tau_parity}. 
Under this coarse-graining, the original sum over one set of the triangular plaquettes (say the up-triangles from Eq.~\ref{eq:3Kagome_Ham}, $\{\triangle^\mu\}$) maps to:
\begin{equation}\label{eq:Kagome_tau_parity}
        \sum_{\triangle^{\mu}} \left( \prod_{i \in \triangle^{\mu}} \sigma^\mu_i \right)
    \simeq \sum_I \tau^\mu_I.
\end{equation}

Furthermore, the operator $\sigma^{\overline{\mu}}_i$ 
now from the down-triangles from $\bigtriangledown^{\overline{\mu}}$ plaquette when acting on a spin within an up-triangular $\triangle^\mu$ plaquette flips the parity of the corresponding $\triangle^\mu$ plaquette. Thus, analogous to Eq.~\ref{eq:tau_parity_flip}, we can define:
\begin{equation}\label{eq:Kagome_tau_parity_flip}
    \sum_{\bigtriangledown^{\overline{\mu}}} 
    \left( \prod_{i \in \bigtriangledown^{\overline{\mu}}}  \sigma^{\overline{\mu}}_i \right)
    \simeq \sum_{\langle IJK \rangle} \tau^{\overline{\mu}}_I \tau^{\overline{\mu}}_J 
    \tau^{\overline{\mu}}_K, 
\end{equation}
where $\langle IJK \rangle$ denotes the $\triangle^{\overline{\mu}}$ plaquettes of the block-decimated
lattice as also indicated in Fig.~\ref{fig:block_decimation_Kagome}.
Thus the Hamiltonian of Eq.~\ref{eq:3Kagome_Ham} maps under the above block-decimation to a Hamiltonian like the Xu-Moore model~\cite{Xu_Moore_PRL_2004} but now defined on triangular lattice:
\begin{equation}
    H \rightarrow H_{bd} = J_\mu \sum_I \tau^\mu_I + 
    J_{\overline{\mu}} \sum_{\langle IJK \rangle} \tau^{\overline{\mu}}_I \tau^{\overline{\mu}}_J 
    \tau^{\overline{\mu}}_K. 
    \label{eq:Kagome_block_decimated_hamil}
\end{equation}
It is important to note that the sign of $J_\mu$ (or $J_{\overline{\mu}}$) does not affect the block decimation procedure. This is because, for an individual triangular plaquette term, both signs of the coupling  yield four degenerate ground states -- i.e., in $\mu$ basis: $\{+++,~ +--,~ -+-,~ --+\}$ for $J_{\mu} <0$ and $\{---,~ -++,~ +-+,~ ++-\}$  for $J_{\mu} >0$) -- and vice versa for the four degenerate excited states.

This block-decimated model on the triangular lattice (Eq.~\ref{eq:Kagome_block_decimated_hamil}) is qualitatively different from the Xu-Moore model on the square lattice (Eq.~\ref{eq:block_decimated_hamil}). 
Unlike the square lattice case, Eq.~\ref{eq:Kagome_block_decimated_hamil} lacks sliding symmetries  due to the triangular nature of the (block-decimated) lattice. 
The ground state of this block-decimated effective Hamiltonian thus exhibits long-range order as shown in Fig.~\ref{fig:Kagome_LRO}. 
Note the order can be either of ferromagnetic type \emph{or} 3-sublattice type with \emph{both} types of ordering being degenerate minima with long-range order for $\frac{J_{\overline{\mu}}}{J_\mu} \gg 1$. 
For a counterpoint to this, see the end of Sec.~\ref{sec:2spin_cases}.
There will also clearly be a self-dual point at $J_\mu = J_{\overline{\mu}}$ in line with Sec.~\ref{subsec:emergence_of_xu_moore}.

\subsection{Pyrocholore lattice}
\label{subsec:pyrochlore}


In this section we consider the pyrocholore lattice of corner-sharing tetrahedra and thus the first three-dimensional lattice in this work. 
The fact that an \guillemotleft anticommuting\guillemotright~$\mathbb{Z}_2$ structure can be naturally accommodated on the pyrocholore lattice was already pointed out in the conclusion section of Ref.~\cite{GSentropy_scipost}. 
Here analogous to Eq.~\ref{eq:4spin_square_hamil}, we will again study a 4-spin \guillemotleft ac\guillemotright~$\mathbb{Z}_2$ QSL which will have features quite similar to the square lattice 4-spin \guillemotleft ac\guillemotright~$\mathbb{Z}_2$ QSL. This is in spite of the supporting lattice being non-bipartite with geometrically frustrated triangular motifs similar to the kagome lattice. 
It may be written as follows:
\begin{equation}
    H =\; J_x \sum_{T_x} \left( \prod_{i \in\; T_x} 
     \sigma^x_i \right)
    +
    J_z \sum_{T_z} \left( \prod_{j \in\; T_z} 
    \sigma^z_j \right),
    \label{eq:4spin_pyrochlore_hamil}
\end{equation}
where $T_x$ and $T_z$ refer to ``up''-tetrahedra and ``down''-tetrahedra as shown in Fig.~\ref{fig:pyro_loop}(a) (or vice-versa). 
The results that follow in the rest of the section constitute exact statements on a three-dimensional quantum spin liquid which is noteworthy and form a distinct variety within the family of $\mathbb{Z}_2$ quantum spin ices. 
It is well-motivated to ask if the above 4-spin model of Eq.~\ref{eq:4spin_pyrochlore_hamil} can be effectively generated using perturbation theory starting with a 2-spin \guillemotleft anticommuting\guillemotright~$\mathbb{Z}_2$ QSL in presence of appropriately chosen transverse field terms or some other mechanics, analogous to how the Kitaev toric code emerges from the Kitaev honeycomb model in one of its perturbative limits. This is left to the future.

\begin{figure*}
    \centering
    \includegraphics[width=0.8\linewidth]{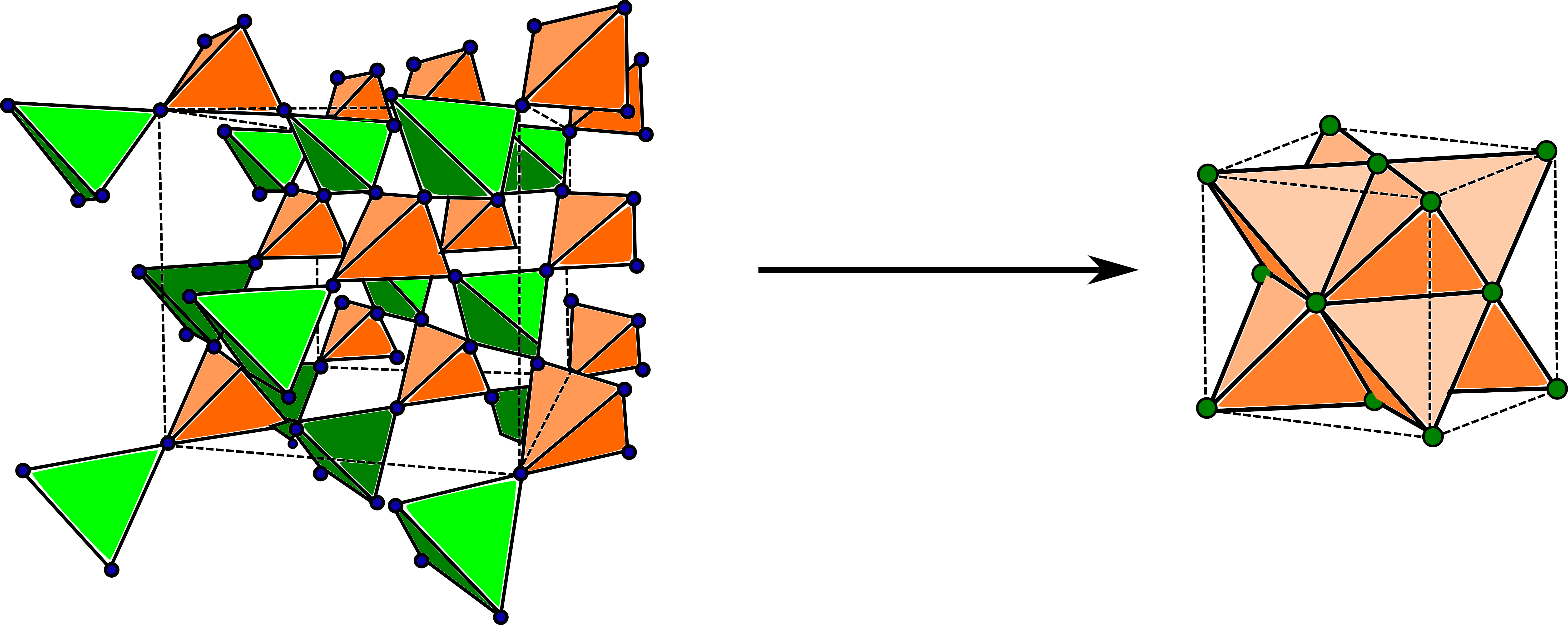}
    \caption{Illustration of the block decimation step for the pyrochlore 4-spin \guillemotleft ac\guillemotright-$\mathbb{Z}_2$ QSL as 
    described in the text of Sec.~\ref{subsec:pyrochlore}}
    \label{fig:block_decimation_pyrochlore}
\end{figure*}

Note that for the 4-spin terms in the Hamiltonian above, the sign of the coupling is not very relevant since the local spectral structure on a single tetrahedron is again just two sets of eight states indexed by the tetrahedral parities analogous to the plaquette parities in the square lattice case. 
The 2-spin form on a single tetrahedron $\sum_{i,j \in T} \sigma^\mu_i \sigma^\mu_j = \left(\sum_{i \in T} \sigma^\mu_i \right)^2 + \text{ a }c$-number is sensitive to the sign of the coupling including the degeneracies for its local spectral structure. 
This point will be discussed further in Sec.~\ref{sec:2spin_cases} and will certainly play an important role in determining the macroscopic properties of the corresponding 2-spin \guillemotleft anticommuting\guillemotright~$\mathbb{Z}_2$ QSLs. 
Here we continue to focus on the 4-spin variant as has been done throughout the work in the previous sections. 
The 2-spin \guillemotleft anticommuting\guillemotright~$\mathbb{Z}_2$ QSLs with either signs of the coupling on the pyrochlore lattice merit their own study.

As a consequence of the 4-spin terms in the Hamiltonian Eq.~\ref{eq:4spin_pyrochlore_hamil}, the model exhibits a ``plain vanilla'' \guillemotleft anticommuting\guillemotright~structure with the following local conserved $\mathbb{Z}_2$ parities on the respective tetrahedra
\begin{eqnarray}
    \sigma^z_i \sigma^z_j \sigma^z_k \sigma^z_l \text{ on the $T_x$ tetrahedra,}
    \label{eq:4spin_pyro_conserved_charge1} \\
    \sigma^x_i \sigma^x_j \sigma^x_k \sigma^x_l \text{ on the $T_z$ tetrahedra.}
    \label{eq:4spin_pyro_conserved_charge2}
\end{eqnarray}
In addition to these, Eq.~\ref{eq:4spin_pyrochlore_hamil} also possesses following conserved quantities defined on the ``missing'' or ``unoccupied'' hexagonal plaquettes as sketched in Fig.~\ref{fig:pyro_loop}(b),
\begin{eqnarray}
    \sigma^x_i \sigma^x_j \sigma^x_k \sigma^x_l \sigma^x_m \sigma^x_n \text{ on the $\hexagon$ plaquettes,}
    \label{eq:6spin_pyro_conserved_charge1} \\
    \sigma^y_i \sigma^y_j \sigma^y_k \sigma^y_l \sigma^y_m \sigma^y_n \text{ on the $\hexagon$ plaquettes,}
    \label{eq:6spin_pyro_conserved_charge2} \\
    \sigma^z_i \sigma^z_j \sigma^z_k \sigma^z_l \sigma^z_m \sigma^z_n \text{ on the $\hexagon$ plaquettes.}
    \label{eq:6spin_pyro_conserved_charge3}
\end{eqnarray}
In the Kitaev representation, Eq.~\ref{eq:4spin_pyrochlore_hamil} can be written as 
\begin{align}\label{eq:Pyro_Majorana_HAM}
    \Tilde{H} =\;  J_x \sum_{T_x}  \left( \prod_{i \in\; T_x} 
     i \gamma^0_i \gamma^x_i \right) 
    +
     J_z \sum_{T_z} \left( \prod_{i \in\; T_z} 
     i \gamma^0_i \gamma^z_i \right).
\end{align}
This model also has the additional
mutually commuting quantities on the tetrahedral motifs 
\begin{eqnarray}
    \left( \gamma^x_i \gamma^x_j \gamma^x_k \gamma^x_l
    \right) \text{ on the $T_x$ plaquettes,}
    \label{eq:4majorana_pyro_conserved_charge3} \\
    \left( \gamma^z_i \gamma^z_j \gamma^z_k \gamma^z_l
    \right) \text{ on the $T_z$ plaquettes,}
    \label{eq:4majorana_pyro_conserved_charge4}
\end{eqnarray}
and also on the hexagonal plaquettes
\begin{eqnarray}
    \prod_{i \in \hexagon} \gamma^{a}, ~~~~~a = \{o,x,y,z\}.
    \label{eq:6majorana_pyro_conserved_charges}
\end{eqnarray}
These operators (Eqs.~\ref{eq:4majorana_pyro_conserved_charge3}–\ref{eq:6majorana_pyro_conserved_charges}) represent symmetries in the extended Hilbert space of the Kitaev Majorana representation, and can be used to reduce the Hamiltonian to an interacting 4-Majorana form similar to the square lattice case
\begin{align}
    \Tilde{H} =\;  J_x \sum_{T_x} \boxed{u_{T_x}} \left( \prod_{i \in\; T_x} 
     \gamma^0_i \right) 
    \; + \;
     J_z \sum_{T_z} \boxed{u_{T_z}} \left( \prod_{i \in\; T_z} 
     \gamma^0_i \right) 
\end{align}
and thus the considerations discussed in Sec.~\ref{sec:hidden_charges} also apply here.

\begin{figure*}
    \centering
    \includegraphics[width=0.7\linewidth]{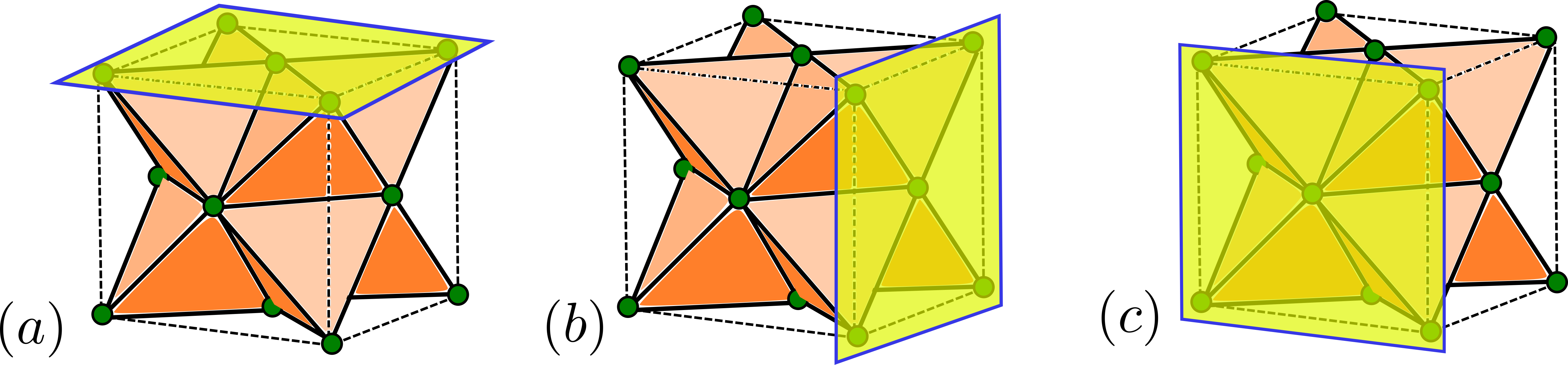}
    \caption{Illustration of the sub-extensive number of the conserved operators defined on entire planes on the block decimated pyrochlore lattice. See Sec.~\ref{subsec:pyrochlore} for details.}
    \label{fig:block_decimation_pyrochlore_loops}
\end{figure*}

In addition -- as might be by now expected for the multi-spin \guillemotleft anticommuting\guillemotright~variants -- there exist two-spin and two-Majorana local symmetry operators of the type discussed in Eqs.~\ref{eq:2spin_symm1}–\ref{eq:2spin_symm4}. 
These take the following forms:
\begin{eqnarray}
    &\sigma_i^x \sigma_j^x~~~\text{for}~i,j \in T_z, \label{eq:2spin_Pyro_symm1}\\
    &\sigma_i^z \sigma_j^z~~~\text{for}~i,j \in T_x,  \label{eq:2spin_Pyro_symm2}
\end{eqnarray}
and corresponding ones in the Majorana representation (Eq.~\ref{eq:Pyro_Majorana_HAM}):
\begin{eqnarray}
    &\gamma_i^x \gamma_j^x~~~\text{for}~i,j \in T_x, \label{eq:2spin_Pyro_symm3}\\
    &\gamma_i^z \gamma_j^z~~~\text{for}~i,j \in T_z.  \label{eq:2spin_Pyro_symm4}
\end{eqnarray}
The above shows the existence of local 2-spin symmetries in a three-dimensional setting of the pyrochlore lattice, analogous to the square (and kagome) lattice case discussed earlier. 
They will also affect the extensive degeneracy counting because of the enlarged \guillemotleft anticommuting\guillemotright~algebra.

We can redo the one-time block-decimation procedure on the pyrochlore lattice Hamiltonian given in Eq.~\ref{eq:4spin_pyrochlore_hamil}. We now define a tetrahedral-parity variable analogous to Eq.~\ref{eq:tau_parity}. Under this coarse-graining, the original sum over on of the sets of up-tetrahedral or down-tetrehedra units -- say $T_\mu$ --  can be mapped to an effective two-level Hamiltonian:
\begin{equation}\label{eq:pyro_tau_parity}
        \sum_{T^{\mu}} \left( \prod_{i \in T^{\mu}} \sigma^\mu_i \right)
    \simeq \sum_I \tau^\mu_I.
\end{equation}

Note that these block-spin tetrahedral parities are defined on a face-centred cubic (FCC) lattice as shown in the right panel of Fig.~\ref{fig:block_decimation_pyrochlore}. Furthermore, the operator $\sigma^{\overline{\mu}}_i$ from $T^{\overline{\mu}}$ tetrahedron when acting on a spin within a  $T^\mu$ tetrahedron flips the parity of the corresponding $T^\mu$ tetrahedron. Thus, analogous to Eq.~\ref{eq:tau_parity_flip}, we now write:
\begin{equation}\label{eq:pyro_tau_parity_flip}
    \sum_{T^{\overline{\mu}}} 
    \left( \prod_{i \in T^{\overline{\mu}}}  \sigma^{\overline{\mu}}_i \right)
    \simeq \sum_{\langle IJKL \rangle} \tau^{\overline{\mu}}_I \tau^{\overline{\mu}}_J 
    \tau^{\overline{\mu}}_K
    \tau^{\overline{\mu}}_L, 
\end{equation}
where $\langle IJKL \rangle$ denotes the 
tetrahedral motifs of the block-decimated FCC lattice as also indicated in Fig.~\ref{fig:block_decimation_pyrochlore}. Thus the Hamiltonian of Eq.~\ref{eq:4spin_pyrochlore_hamil} maps under the
above block-decimation to a Xu-Moore like model \cite{Xu_Moore_PRL_2004} defined on a FCC lattice as a network of edge-sharing tetrahedra (Fig.~\ref{fig:block_decimation_pyrochlore}):
\begin{equation}
    H \rightarrow H_{bd} = J_\mu \sum_I \tau^\mu_I + 
    J_{\overline{\mu}} \sum_{\langle IJKL \rangle} \tau^{\overline{\mu}}_I \tau^{\overline{\mu}}_J 
    \tau^{\overline{\mu}}_K 
    \tau^{\overline{\mu}}_L. 
    \label{eq:pyro_block_decimated_hamil}
\end{equation}
The block decimated Hamiltonian $H_{bd}$ again exhibits a sub-extensive number of non-local conserved operators and thus is of the Xu-Moore type, analogous to the square lattice case discussed in Sec.~\ref{subsec:emergence_of_xu_moore} and unlike the Kagome case discussed in the previous Sec.~\ref{subsec:kagome}. 
The non-local conserved operators are products of $ \tau^\mu $ operators defined on \emph{entire planes} that intersect the pyrochlore tetrahedra in such a way that two sites of each tetrahedron lie on the plane as illustrated in Fig.~\ref{fig:block_decimation_pyrochlore_loops}. 
For example the planes could be the family $(100)$ or $(110)$, with the $(100)$ being the natural or primary choice.
Thus Xu-Moore physics will be applicable for the 4-spin pyrochlore \guillemotleft ac\guillemotright-$\mathbb{Z}_2$ QSL in terms of the tetrahedral parities. 
This is apart from quantum spin liquidity in the underlying spin-$\frac{1}{2}$ $\{\sigma^\mu_i\}$ variables stipulated by the \guillemotleft anticommuting\guillemotright-$\mathbb{Z}_2$ algebra.
Here Xu-Moore physics will concern spontaneously broken or unbroken sub-system symmetries on the entire sets of planes along with corresponding unconventional non-local bond orders and self-dualities.

We will end the non-bipartite discussion by pointing out the topological degeneracies from global superselection sector analysis for the present case of the 4-spin pyrochlore \guillemotleft ac\guillemotright-$\mathbb{Z}_2$ QSL. 
The arguments essentially are a repeat of those encountered before for the square lattice case (Eq.~\ref{eq:4spin_square_hamil} and Sec.~\ref{subsec:superselection_4spin_square},~\ref{sec:2spin_cases}) and the kagome lattice case (Eq.~\ref{eq:3Kagome_Ham} and towards the end of Sec.~\ref{subsec:kagome}) and are repeated here for completeness. 
For Eq.~\ref{eq:4spin_pyrochlore_hamil}, there exist non-local loop operators quite like those in Fig.~\ref{fig:corner_sq_superselection} in Sec.~\ref{subsec:superselection_4spin_square} that remain conserved and would thread across periodic boundary conditions. 
These non-local loop operators are basically products of either $\sigma^x$ or $\sigma^z$ operators along non-contractible loops threading through the three-dimensional lattice, such that each loop encounters exactly two lattice sites on every tetrahedron as it traverses the pyrochlore lattice. 
These operators can again be decomposed in terms of the local 2-spin conserved charges of Eqs.~\ref{eq:2spin_Pyro_symm1},~\ref{eq:2spin_Pyro_symm2}.

As before, one can define more general models following the discussion of Sec.~\ref{sec:2spin_conservation},
\begin{equation}\label{eq:Big_Pyro_4spin_model}
    H_{\text{gen1}} =  
    \sum_{\mu \in {R,G}} \sum_{T_\mu} \left( \alpha^\mu \prod_{i \in\; T_\mu} 
   \sigma^x_i + \beta^\mu \prod_{i \in\; T_\mu} 
   \sigma^z_i \right)
\end{equation}
or
\begin{equation}\label{eq:Big1_Pyro_4spin_model}
    H_{\text{gen2}} =  
    \sum_{\mu \in {x,z}} \sum_{T_\mu} J^\mu \left(  \prod_{i \in\; T_\mu} 
   \sigma^\mu_i \right)
   +
   \sum_{\hexagon} J^y \left( \prod_{i \in\; \hexagon} 
   \sigma^y_i \right)
\end{equation}
or a combination for which such 2-spin local conserved charges are absent. 
From a model construction point of view, Eq.~\ref{eq:Big_Pyro_4spin_model} is perhaps slightly less unwieldy than Eq.~\ref{eq:Big1_4spin_model} since the tetrahedral motifs and the hexagonal motifs are different geometric objects with different co-dimensions inside the pyrochlore lattice.
The above generalizations void the aforesaid 2-spin decomposition of the non-local operators that however remain conserved under Eq.~\ref{eq:Big_Pyro_4spin_model} or ~\ref{eq:Big1_Pyro_4spin_model}. 
This mirrors the structure of non-local loop operators in the square and Kagome lattice cases (see Sec.~\ref{sec:superselection} for details). 
As a result, using similar global superselection sector arguments, the model of Eq.~\ref{eq:Big_Pyro_4spin_model} exhibits an eight-fold \emph{topological} ground state degeneracy at the boundary of which sits Eq.~\ref{eq:4spin_pyrochlore_hamil} when $\alpha_x = \beta_z = 0$ or $\alpha_z = \beta_x = 0$.
One again expects to generate Eq.~\ref{eq:Big1_Pyro_4spin_model} from
\begin{equation}\label{eq:Big2_Pyro_4spin_model}
    H_{\text{gen3}} =  
    \sum_{\mu \in {x,z}} \sum_{T_\mu} J^\mu \left(  \prod_{i \in\; T_\mu} 
   \sigma^\mu_i \right)
   +
   J^y \sum_i \sigma^y_i 
\end{equation}
perturbatively in the limit $\frac{J^y}{J^\mu} \ll 1$.
The pyrochlore \guillemotleft ac\guillemotright-$\mathbb{Z}_2$ QSL construction of this section hence provides a concrete example of a three-dimensional topological quantum spin liquid! 
This coexists with unconventional Xu-Moore many-body order in the one-time block-decimated tetrahedra parity variables.

\section{Block decimation considerations for 
\guillemotleft anticommuting\guillemotright~$\mathbb{Z}_2$ quantum spin liquids with 2-spin couplings}
\label{sec:2spin_cases}

In the previous 
Secs.~\ref{sec:emergence_of_many_body_order}, ~\ref{sec:superselection} and ~\ref{sec:non_bipartite}, we saw that a lot more information could be gleaned from the 4-spin \guillemotleft ac\guillemotright-$\mathbb{Z}_2$ QSLs compared to the 2-spin case discussed previously only in Sec.~\ref{sec:hidden_charges}.
For the 2-spin \guillemotleft ac\guillemotright-$\mathbb{Z}_2$ QSLs, it is harder to glean more information apart from the exact statements contained in Ref.~\cite{GSentropy_preprint} and Sec.~\ref{sec:hidden_charges}. 
In this section, we will see what the one-time block decimation has to say about these 2-spin \guillemotleft ac\guillemotright-$\mathbb{Z}_2$ QSLs.
The resultant effective Hamiltonian will clearly have at most bilinear coupling between the effective one-time block decimated degrees of freedom. 
For the 2-spin \guillemotleft ac\guillemotright-$\mathbb{Z}_2$ QSL of Eq.~\ref{eq:2spin_square_hamil} or its analog on the pyrochlore lattice, let us again define a block decimation based on the local spectral structure of the elementary plaquette motifs.
I.e., $\left\{\boxed{x}\right\}$ or $\left\{\boxed{z}\right\}$ for the square plaquettes on the square lattice, and $\{T_x\}$ or $\{T_z\}$ for the tetrahedral motifs on the pyrochlore lattice respectively. 
For the 4-spin \guillemotleft ac\guillemotright-QSL of Eq.~\ref{eq:4spin_square_hamil} it had led to two-valued variables $\{\tau^\mu_I\}$ (cf. Eq.~\ref{eq:tau_parity}) on the decimated lattice as shown in Fig.~\ref{fig:block_decimation}. 
For the 2-spin case, the degeneracies of the local plaquette spectral structure will be sensitive to the form of the local plaquette Hamiltonian, $\pm \sum_{\langle i,j \rangle \in \boxed{\mu}} \sigma^\mu_i \sigma^\mu_j$ or $\pm \sum_{i,j \in \boxed{\mu}} \sigma^\mu_i \sigma^\mu_j$, etc. 
However, generically speaking, in cases with uniform couplings on the plaquettes, the local plaquette energies take three values $\pm 4 J_\mu$ and $0$. 
Thus upon coarse graining we will arrive at a three-valued variable, which for the case of Eq.~\ref{eq:2spin_square_hamil} in analogy with Eq.~\ref{eq:tau_parity} can be written as 
\begin{equation}
 4   \begin{pmatrix}
1 & 0 & 0\\
0 & 0 & 0\\
0 & 0 & -1
\end{pmatrix}
    = \tilde{\tau}^\mu_I \simeq \sum_{\langle i,j \rangle \in \boxed{\mu}}
    \sigma^\mu_i \sigma^\mu_j
    \label{eq:tau_2spin}
\end{equation}
with the block decimated lattice again resulting in a square lattice as in Fig.~\ref{fig:block_decimation} (and the FCC lattice for the pyrochlore case as in Fig.~\ref{fig:block_decimation_pyrochlore}). Now we ask ourselves what is the effect of inter-plaquette 2-spin couplings $\sigma^{\bar{\mu}}_i \sigma^{\bar{\mu}}_j$ after block decimation in the 2-spin case? In other words, what is the analog of the $\tau^{\bar{\mu}}_I$ operator of Sec.~\ref{subsec:emergence_of_xu_moore} now? It will clearly possess a $3\times3$ matrix representation. Since $\sigma^{\bar{\mu}}$ will lead to spin flips for the underlying $\sigma^{\mu}$ variable, it will lead to a raising/lowering type of operator action for the effective block-decimated three-valued plaquette variable $\tau^\mu_I$. We recall from Sec.~\ref{subsec:emergence_of_xu_moore} for the 4-spin case of Eq.~\ref{eq:4spin_square_hamil}, this had led to a ``bit-flip'' action in the two-valued plaquette parity $\tau^\mu_I$ and thus was naturally ascribed to as $\tau^{\bar{\mu}}_I$. In the 2-spin case, a similar step will lead to the general form
\begin{equation}
    \tilde{\tau}^{\bar{\mu}}_I = 
    \begin{pmatrix}
0 & a & 0\\
a & 0 & b\\
0 & b & 0
\end{pmatrix},
\end{equation}
where $a$,$b$ represent the net effect of all the quantum mechanical processes and quantum interferences that go into the raising/lowering action discussed earlier. 
The actual values of $a$,$b$ will be sensitive to, for example, the degeneracy pattern of the underlying local plaquette spectral structure, etc. 
The resultant 2-spin couplings will take the form $J_{\bar{\mu}} \tilde{\tau}^{\bar{\mu}}_I \tilde{\tau}^{\bar{\mu}}_J$ on all the bonds that survive the block decimation. For the case of Eq.~\ref{eq:2spin_square_hamil}, this will be the nearest neighbour bonds $\langle I,J \rangle$ of the block decimated lattice (Fig.~\ref{fig:block_decimation}). 
This raising/lowering action will be accompanied by a change of the local plaquette energy on the ($\left\{ \boxed{\mu} \right\} or \{T_\mu\}$) plaquettes with which we set up the block decimation step. 
Thus the resultant effective block-decimated Hamiltonian for Eq.~\ref{eq:2spin_square_hamil} will be
\begin{equation}
    H \rightarrow H_{bd} = 
    J_\mu \sum_I \tilde{\tau}^\mu_I
    +
    J_{\bar{\mu}} \sum_{\langle I,J \rangle} 
    \tilde{\tau}^{\bar{\mu}}_I \tilde{\tau}^{\bar{\mu}}_J,
    \label{eq:block_decimated_2spin_hamil}
\end{equation}
which is like a quantum Ising model but now built out
of the three-valued local plaquette energy variables.
One needs to study and analyze the above Hamiltonian
to learn more about the universal physics of 2-spin 
\guillemotleft ac\guillemotright-$\mathbb{Z}_2$ QSLs. 
Whatever that universal physics may be, it will coexist with the extensive residual ground state entropy and quantum spin liquidity in the original spin-$\frac{1}{2}$ variables $\sigma^\mu_i$.

We can already say that the above Hamiltonian in Eq.~\ref{eq:block_decimated_2spin_hamil} with three-level degrees of freedom will have a self-duality at $J_\mu = J_{\bar{\mu}}$ with appropriate operator duality relations. 
This is essentially in analogy with the self-duality of the the Xu-Moore model in Eq.~\ref{eq:block_decimated_hamil} that becomes self-evident when looking at the underlying 4-spin \guillemotleft ac\guillemotright-$\mathbb{Z}_2$ QSL of Eq.~\ref{eq:4spin_square_hamil} while setting up the block decimation as in Sec.~\ref{subsec:block_decimation}. 
It may be possible that the block decimation arguments given in this section may be used give a new perspective on other microscopic operator level dualities existing already in the literature, i.e., self-dualities of Hamiltonian that are not that obvious in the original variables may become transparent and obvious in terms of some underlying Hamiltonian and degrees of freedom.

Redoing the above for the case of the Kagome lattice with its corner-sharing triangle property, the local plaquette structure will now give a two-valued block variable quite like in Sec.~\ref{subsec:block_decimation}! 
For the cases with uniform couplings on the plaquettes, the local plaquette energies take two values $\{3J_\mu,-J_\mu\}$ that can be mapped to $\pm J_\mu$ through additive constants, and thus we now write
\begin{equation}
    \begin{pmatrix}
1 & 0 \\
0 & -1
\end{pmatrix}
    = \tilde{\tau}^\mu_I \simeq \sum_{\langle i,j \rangle \in \bigtriangleup^\mu}
    \frac{\left( \sigma^\mu_i \sigma^\mu_j -  1\right)}{2}
    \label{eq:tau_2spin_kagome}
\end{equation}
for the one-time decimation block variable on the plaquettes $\{\bigtriangleup^\mu\}$, and
\begin{equation}
    \tilde{\tau}^{\bar{\mu}}_I = 
    \begin{pmatrix}
0 & 1 \\
1 & 0
\end{pmatrix}
\end{equation}
for the raising/lowering action that will take place due to the inter-plaquette 2-spin couplings. Thus we will arrive at the following effective block-decimated Hamiltonian 
\begin{equation}
    H \rightarrow H_{bd} = 
    J_\mu \sum_I \tilde{\tau}^\mu_I
    +
    J_{\bar{\mu}} \sum_{\langle I,J \rangle} 
    \tilde{\tau}^{\bar{\mu}}_I \tilde{\tau}^{\bar{\mu}}_J,
    \label{eq:block_decimated_2spin_kagome}
\end{equation}
with two-valued block variables that track the local plaquette energies on the block decimated plaquettes. 
The above Hamiltonian lives on a triangular lattice after the block decimation similar to Fig.~\ref{fig:block_decimation_Kagome} but now for the 2-spin \guillemotleft ac\guillemotright-$\mathbb{Z}_2$ QSL on the kagome lattice with $120^\circ$ rotational symmetry around the plaquette centers. 
Note that the above effective Hamiltonian Eq.~\ref{eq:block_decimated_2spin_kagome} is essentially the quantum Ising model on the triangular lattice! 
The nature of long-range order will be dictated by the sign of $J_{\overline{\mu}}$, i.e., ferromagnetic order for $J_{\overline{\mu}} < 0$  as sketched in the left of Fig.~\ref{fig:Kagome_LRO}, and three-sublattice order for $J_{\overline{\mu}} > 0$ as sketched in the right of Fig.~\ref{fig:Kagome_LRO}.  
This also implies that the quantum phase transition between the phases  for $\frac{J_{\overline{\mu}}}{J_\mu} \gg 1$ and $\frac{J_{\overline{\mu}}}{J_\mu} \ll 1$ with antiferromagnetic signs $J_\mu>0$, $J_{\overline{\mu}}>0$ will be described by Damle's theory for 3-sublattice order melting transitions given as laid out in Refs.~\cite{Damle_prl_2015,Rakala_etal_arxiv_2021}.

\section{Conclusion}
\label{sec:conclu}

In summary, this work studies several model constructions with spin-$\frac{1}{2}$ or qubit degrees of freedom on various lattices that show quantum spin liquidity, residual ground state entropies, and for the multi-spin coupling variants, topological degeneracies and unconventional many-body order of the Xu-Moore type~\cite{Xu_Moore_PRL_2004}. 
Quantum spin liquidity and residual entropy results follow directly based on Ref.~\cite{GSentropy_scipost} and are undergirded by the \guillemotleft anticommuting\guillemotright~algebra of local $\mathbb{Z}_2$ charges. 
The results on topological degeneracies and unconventional many-body order are developed in Secs.~\ref{sec:emergence_of_many_body_order},~\ref{sec:superselection},~\ref{sec:non_bipartite}. 
This work also exposes in Sec.~\ref{sec:hidden_charges} the non-standard nature of these models when viewed as lattice gauge theories with a fermionic Majorana matter sector and a $\mathbb{Z}_2$ gauge sector. 
This becomes apparent in the Kitaev representation of these QSL models and more may be learned by taking this perspective forward. 

These models have thus been named here as \guillemotleft anticommuting\guillemotright~$\mathbb{Z}_2$ quantum spin liquids (\guillemotleft ac\guillemotright-$\mathbb{Z}_2$ QSLs in short) following the nomenclature laid down in Ref.~\cite{GSentropy_scipost}. 
In this paper we restricted ourselves to translationally invariant and nearest-neighbour form of the couplings for simplicity. 
Many of the results such as those regarding quantum spin liquidity and topological degeneracies extend to more general situations such as the absence of translational symmetry for disordered coupling strengths, etc.  
It is worth emphasizing that the pyrochlore lattice \guillemotleft ac\guillemotright-$\mathbb{Z}_2$ QSL studied in Sec.~\ref{subsec:pyrochlore} and Sec.~\ref{sec:2spin_cases} provide constructive models with provable quantum spin liquidity in three dimensions. 
QSL variants on other three dimensional lattices such as the cubic lattice or the non-bipartite hyperkagome lattice can also be written down.

Concerning the site-interlinked geometrical property of the local conserved charges that leads to their \guillemotleft anticommuting\guillemotright~algebra, one may wonder whether it is limited to the $\mathbb{Z}_2$ group or can be extended to the $\mathbb{U}(1)$ or $\mathbb{SU}(2)$ groups. 
With spin-$\frac{1}{2}$ qubits, this appears to be not naturally possible. 
E.g. take a ``locally $\mathbb{U}(1)$-symmetric'' bond term of the form $\sigma^x_i \sigma^x_i + \sigma^y_i \sigma^y_j$ on the bond $\langle i,j \rangle$. 
This bond has the associated  quantity $(\sigma^z_i+\sigma^z_j)$ that is conserved under the locally $\mathbb{U}(1)$-symmetric bond term, i.e., $[(\sigma^z_i+\sigma^z_j),\sigma^x_i \sigma^x_i + \sigma^y_i \sigma^y_j]=0$. 
On an adjoining bond $\langle j,k \rangle$, we may take the natural possibility of a different locally $\mathbb{U}(1)$-symmetric bond term $\sigma^y_j \sigma^y_k + \sigma^z_j \sigma^y_k$. 
It comes similarly with the associated quantity $(\sigma^x_j+\sigma^x_k)$ which satisfies $[(\sigma^x_j+\sigma^x_k),\sigma^y_j \sigma^y_k + \sigma^z_j \sigma^y_k]=0$. 
However, $\{(\sigma^z_i+\sigma^z_j),(\sigma^x_j+\sigma^x_k)\}\neq0$ as well as $[(\sigma^z_i+\sigma^z_j),\sigma^x_i \sigma^x_i + \sigma^y_i \sigma^y_j + \sigma^y_j \sigma^y_k + \sigma^z_j \sigma^y_k]\neq0$ and $[(\sigma^x_j+\sigma^x_k),\sigma^x_i \sigma^x_i + \sigma^y_i \sigma^y_j + \sigma^y_j \sigma^y_k + \sigma^z_j \sigma^y_k]\neq0$. 
This shows that an \guillemotleft anticommuting\guillemotright-$\mathbb{U}(1)$ structure is not natural. 
With higher spins or qualitatively other degrees of freedom (e.g. the various representations of the $\mathbb{SU}(N)$ group), there may be some possibilities. 
But it might well be the case that such structures are restricted to discrete groups such as $\mathbb{Z}_N$ (or perhaps other acyclic discrete groups not necessarily reducible to direct sums of cyclic groups). 
One may also ask if one can find a lattice geometry or a model that totally avoids the \guillemotleft anticommuting\guillemotright~structure in a much stronger way, i.e., there do not exist any local \guillemotleft anticommuting\guillemotright~algebras. 
This possibility is discussed briefly in Appendix~\ref{app:no_hidden_charges} since this curiosity is not the focus of this paper.

The multi-spin models of Eq.~\ref{eq:4spin_square_hamil},~\ref{eq:3Kagome_Ham} and ~\ref{eq:4spin_pyrochlore_hamil} have quite a family resemblance to the Kitaev toric code and other related Hamiltonians. 
However, their physics is quite different due to the \guillemotleft anticommuting\guillemotright~algebraic structure that is not part of the toric code and the Kitaev honeycomb model and more generally the Levin-Wen models. 
Thus with respect to the topological degeneracies found in Eq.~\ref{eq:4spin_square_hamil},~\ref{eq:3Kagome_Ham} and ~\ref{eq:4spin_pyrochlore_hamil} which are limits of the more general Hamiltonians in Eq.~\ref{eq:Big_4spin_model},~\ref{eq:Big_Kag_3spin_model} and ~\ref{eq:Big_Pyro_4spin_model} respectively, it remains to be seen how robust is the  topological order. 
I.e., what would happen to the topological degeneracy under generic perturbations such as a ``transverse'' field? 
Note a magnetic field in any direction will generically spoil any \guillemotleft anticommuting\guillemotright~algebraic structure. 
We conjecture here that it indeed is robust up till some value of such perturbations. 
In a similar vein, just as there is an elegant loop superposition picture for the wavefunctions of the Kitaev toric code and Levin-Wen models more generally, is there an analogous geometric picture not necessarily involving loops per se for the \guillemotleft ac\guillemotright-$\mathbb{Z}_2$ QSLs? 
This may be asked separately or in conjunction for both the multi-spin variants that were the focus of this work and/or the 2-spin variants that can be said to be the primary realizations of \guillemotleft anticommuting\guillemotright~structures.

If the claim above regarding the robustness of the topological degeneracy is true, the topological degeneracy coexisting with Xu-Moore order in plaquette parities and underlying residual entropy in \guillemotleft ac\guillemotright-$\mathbb{Z}_2$ QSLs would then represent a distinct, stable many-body topological phase in the space of spin-$\frac{1}{2}$ Hamiltonians when compared to the many-body topological order of the toric code. 
The coexistence of many-body topological order with the \guillemotleft anticommuting\guillemotright~algebraically generated extensive degeneracies may appear strange, but there is at least one other qualitatively different case where gaplessness has been shown to coexist with (symmetry-protected) topological order. 
This goes by the name of gapless symmetry-protected topological order~\cite{gSPT1_Scafiddi_2017,gSPT2_Thorngren_2021} which has received attention recently (see the recent Ref.~\cite{gSPT_latest_Li_2024} and references therein). 

We also note in passing that Kitaev wrote down the toric code with its mutually commuting local algebras in view of fault tolerance for quantum information processing. 
From this perspective, it remains to be seen if the \guillemotleft ac\guillemotright-$\mathbb{Z}_2$ QSL Hamiltonians as qubit codes have any bearing on fault tolerance in quantum information processing vis-a-vis quantum error correction, especially given their family resemblance to Kitaev toric code and other related surface codes. 
One example in the literature that comes close is Ref.~\cite{Bravyi_Poulin_2013} with a 6-spin coupling for its Hamiltonian terms. 
It has an underlying 3-spin \guillemotleft anticommuting\guillemotright~algebra which was not explored by its authors. This seems to affirm the conjecture made by one of the authors recently that the \guillemotleft ac\guillemotright-$\mathbb{Z}_2$ QSLs may be topological subsystem codes~\cite{Pujari_youtube_2025}. 
Thus the multi-spin \guillemotleft ac\guillemotright-$\mathbb{Z}_2$ QSLs written down in this paper might provide simpler examples of topological subsystem codes and deserve further study within the context of quantum error correction and fault tolerance.

Within the context of strongly correlated quantum matter, at a technical level, we see that the structure of the non-local operators from Sec.~\ref{sec:superselection} on the global superselection sector analysis are not the same for the Kitaev toric code and that for the \guillemotleft  AC\guillemotright~$\mathbb{Z}_2$ QSLs. 
One may imagine a scenario where effective star and plaquette terms emerge at a bigger length scale say after some renormalization procedure. 
In this scenario, the topological order in the multi-spin \guillemotleft ac\guillemotright-$\mathbb{Z}_2$ QSLs will again be of the toric code type. 
However given that the one-step block-decimation leads to the Xu-Moore Hamiltonian with subsystem symmetries, it is hard to imagine how the above scenario can emerge.
This again argues for distinct nature of the topological order in the multi-spin \guillemotleft ac\guillemotright-$\mathbb{Z}_2$ QSLs.

Of course the exact residual ground state entropy will also be lost in presence of generic perturbations that void the \guillemotleft anticommuting\guillemotright~structure. 
It will lead to a very high many-body density of states near the putatively multi-fold topological ground state manifold which would make it effectively like a residual entropy at higher temperatures. 
The topological degeneracies would then be expected to split much slower than other degeneracies such as those coming from the \guillemotleft anticommuting\guillemotright~algebra say from a numerical diagnostic point of view. 
The verification of this conjecture is left to the future.

This work answers some of the open questions posed in Ref.~\cite{GSentropy_scipost} that can be broadly contextualized as different infinite-dimensional representations of the \guillemotleft anticommuting\guillemotright~algebras on the lattice in a representation-theoretic sense~\cite{VShenoy_private}. 
Others questions pertaining to eigenstate thermalization, quantum chaos and continuum field theoretic formulation aspects posed in Ref.~\cite{GSentropy_scipost} remain open. 
As an aside, a continuum field theory for Xu-Moore many-body order was written down in Ref.~\cite{Xu_Moore_PRL_2004}.
We end with the claim that \guillemotleft anticommuting\guillemotright~$\mathbb{Z}_2$ quantum spin liquids discussed here and Ref.~\cite{GSentropy_scipost} appear to open a new vista to a wide playground or garden of novel strongly correlated many-body orders.

\section{Acknowledgements}

Discussions with Abhijeet Gadde, Sumilan Banerjee, Subhro Bhattacharjee, Ankush Chaubey, Kedar Damle, Beno\^ it Dou\c cot, Nisheeta Desai, Souvik Kundu, Gautam Mandal, Shiraz Minwalla, Krushna Chandra Sahoo, Diptiman Sen, Vijay Shenoy, Rajdeep Sensarma, Sthitadhi Roy, Vikram Tripathi and Julien Vidal are gratefully acknowledged. S.P. especially thanks Roderich Moessner for feedback and encouragement. S.P. acknowledges funding support from SERB-DST, India (superseded by ANRF-DST established through an Act of Parliament: ANRF Act, 2023) via Grant No. MTR/2022/000386. H.N. acknowledges the support of the Department of Atomic Energy, Government of India, under project no. RTI4001.

\bibliography{refs}


\newpage
\begin{appendix}

 \section{Inter-convertibility arguments with reference 
 to global superselection sector analyses}
\label{app:interconvertibility}

In Sec.~\ref{sec:superselection}, we mentioned the existence of a certain number of non-local conserved operators for the different models presented that are not ``inter-convertible'' into each other.
The meaning of inter-convertibility will be elaborated here.
This will clarify why only a specific number of such independent operators exists, namely they cannot be transformed or converted into one another through multiplication with local conserved operators that are available in the model under consideration.

We begin with the toric code which features four non-local conserved operators: $\{O^x_h, O^z_h, O^x_v, O^z_v\}$ as illustrated in Fig.~\ref{fig:ktc_superselection}(a). 
Consider $O^x_h$, defined along a horizontal loop passing through the centers of plaquettes. 
There are many such operators; indeed one can define them on every horizontal loop through the plaquette centers, however, not all of these operators are independent.
For instance, take an operator $O^x_h$ defined on a loop $\mathcal{L}_h$. 
By multiplying it with a string of $A_s$ operators, one can shift this operator to a new loop $\mathcal{L}_h \pm 1$. 
This implies that it is possible to move $O^x_h$ upwards or downwards using a string of $A_s$ operators, indicating that all such operators are not independent. 
Similarly, one can shift $O^z_h$ operators by multiplying them with strings of $B_p$ operators. 
The same argument applies to shifting the $O_v^x$ and $O_v^z$ operators left or right by multiplying them respectively with a string of $A_s$ and $B_p$ operators. 
Therefore we can restrict to a single representative from each set of $\{O^x_h\},~ \{O^z_h\},~ \{O^x_v\},~ \{O^z_v\}$. 
However, it is not possible to convert between $O^x_h$, $O^z_h$, $O^x_v$, and $O^z_v$ using local conserved quantities. 
Hence these four operators are independent  or non-interconvertible non-local conserved quantities.

Similarly, for the 4-spin model of Eq.~\ref{eq:Big_4spin_model} on the corner-sharing square lattice, there exist many local conserved operators (given in Eqs.~\ref{eq:4spin_conserved_charge5},~\ref{eq:4spin_conserved_charge7}) using which we can show that there are only four non-interconvertible non-local operators.
Let us begin with $O_h^x$ and $O_h^z$ as shown in Fig.~\ref{fig:BIG_sq_model}. 
This operator can be shifted upward or downward by multiplying it with operators defined in Eq.~\ref{eq:4spin_conserved_charge5} (~\ref{eq:4spin_conserved_charge7}), where the product is taken over spins on an empty or colourless plaquette.
Similarly, $O_v^x$  and $O_v^z$ can be shifted left or right by multiplying respectively with Eq.~\ref{eq:4spin_conserved_charge5} and Eq.~\ref{eq:4spin_conserved_charge7} again defined on empty plaquettes. 
Like the toric code, here too we are left with four independent or non-interconvertible non-local conserved quantities.
For the general three-spin model (Eq.~\ref{eq:Big_Kag_3spin_model}) on the Kagome lattice too, we have four non-local conserved quantities as shown in Fig.~\ref{fig:Kagome_superselection}. To shift any of $O^x_h$, $O^z_h$, $O^x_v$, or $O^z_v$, we can use local conserved quantities $\left(\prod_{i \in \hexagon} \sigma_i^x,\ \prod_{i \in \hexagon} \sigma_i^z\right)$, defined on (empty or colourless) hexagonal plaquettes. As a result, we are again left with four non-interconvertible non-local conserved operators.

\section{Argument for absense of topological degeneracy for model in Eq.~\ref{eq:4spin_square_hamil}}
\label{app:TD_details}

As discussed in Sec.~\ref{sec:2spin_conservation}, the 4-spin model defined in Eq.~\ref{eq:4spin_square_hamil} lacks topological protection of the four-fold degeneracy due to the presence of two-spin conserved quantities of Eqs.~\ref{eq:2spin_symm1},~\ref{eq:2spin_symm2}). The models in Eq.~\ref{eq:Big_4spin_model} or ~\ref{eq:Big1_4spin_model} in contrast do not possess such local two-spin conserved quantities and support topological degeneracy. 
Here we provide a perturbative argument to make this point more explicit following similar arguments in the literature.

Let us consider a ground state $|\psi_1\rangle$ and  its partner $|\psi_2\rangle = O_h^z |\psi_1\rangle$. 
$\psi_2\rangle$ is another degenerate ground state 
following the arguments of Sec.~\ref{sec:superselection} and Tab.~\ref{tab:four_fold_degeneracy} where we have used $O_h^z = \prod_{i \in \mathcal{L}_h} \sigma_i^z$ to generate the partner state here (cf. Fig.~\ref{fig:corner_sq_superselection}). 
We now introduce a perturbation of the form
\begin{equation}
    H_\epsilon = \epsilon \sum_i \sigma_i^z,
\end{equation}
and ask whether there exists a non-zero ``scattering amplitude'' \( A_{12} \) between the two ground states in the thermodynamic limit.

At $n$-th order in perturbation theory, the transition amplitude takes the form:
\begin{align}
    A_{12} &\sim \epsilon^n \langle\psi_1| \left( \sum_i \sigma_i^z \right)^n |\psi_2\rangle \\
           &= \epsilon^n \langle\psi_1| \left( \sum_i \sigma_i^z \right)^n \prod_{i \in \mathcal{L}_h} \sigma_i^z |\psi_1\rangle.
\end{align}
For $n=2$, a representative contribution to the amplitude is:
\begin{align}
    A_{12} &\sim \epsilon^2 \langle\psi_1| \left( \sigma_1^z \sigma_2^z \right) \cdot \left( \sigma_1^z \sigma_2^z \sigma_3^z \sigma_4^z \cdots |_{\mathcal{L}_h} \right) |\psi_1\rangle \\
           &\sim \epsilon^2 \langle\psi_1| \left( \sigma_3^z \sigma_4^z \cdots |_{\mathcal{L}_h} \right) |\psi_1\rangle
           \label{eq:partial_product}.
\end{align}
Now, the presence of local two-spin conserved quantities in the model of Eq.~\ref{eq:4spin_square_hamil} allows such partial products of $\sigma^z$ operators in Eq.~\ref{eq:partial_product} to be expressed in terms of the two-spin symmetries, leading to:
\begin{equation}
    A_{12} \sim \epsilon^2 \neq 0.
\end{equation}
Hence, the two ground states are not protected from mixing by local perturbations, and the model lacks true topological degeneracy.

On the other hand, for the model defined in Eq.~\ref{eq:Big_4spin_model} or ~\ref{eq:Big1_4spin_model}, which lacks such local conserved quantities, a transition from $|\psi_1\rangle$ to $|\psi_2\rangle$ would require a perturbation acting non-trivially on \emph{all} the spins along the non-local loop $\mathcal{L}_h$. The lowest order at which this transition can occur is:
\begin{equation}
    A_{12} \sim \epsilon^{\text{number of sites in }\mathcal{L}_h},
\end{equation}
which vanishes exponentially in the thermodynamic limit ($\text{number of sites in }\mathcal{L}_h \to \infty$) for any $\epsilon \to 0$:
\begin{equation}
    \lim_{\epsilon \rightarrow 0} \; \lim_{|\mathcal{L}_h| \to \infty} \; A_{12} = 0.
\end{equation}
where we take the limits in the order given above.
This is essentially the same argument as that of the topological protection of the four-fold degeneracy of the toric code
given in Sec.~3 of Ref.~\cite{Kitaev_2003}.
This completes the argument presented in Sec.~\ref{sec:superselection} for the presence of topological degeneracy applies to the model in Eq.~\ref{eq:Big_4spin_model}.


\section{Graphs with neither local conserved quantities nor $n$-Majorana form symmetries}
\label{app:no_hidden_charges}

\begin{figure}[t]
    \centering
    \includegraphics[width=0.8\linewidth]{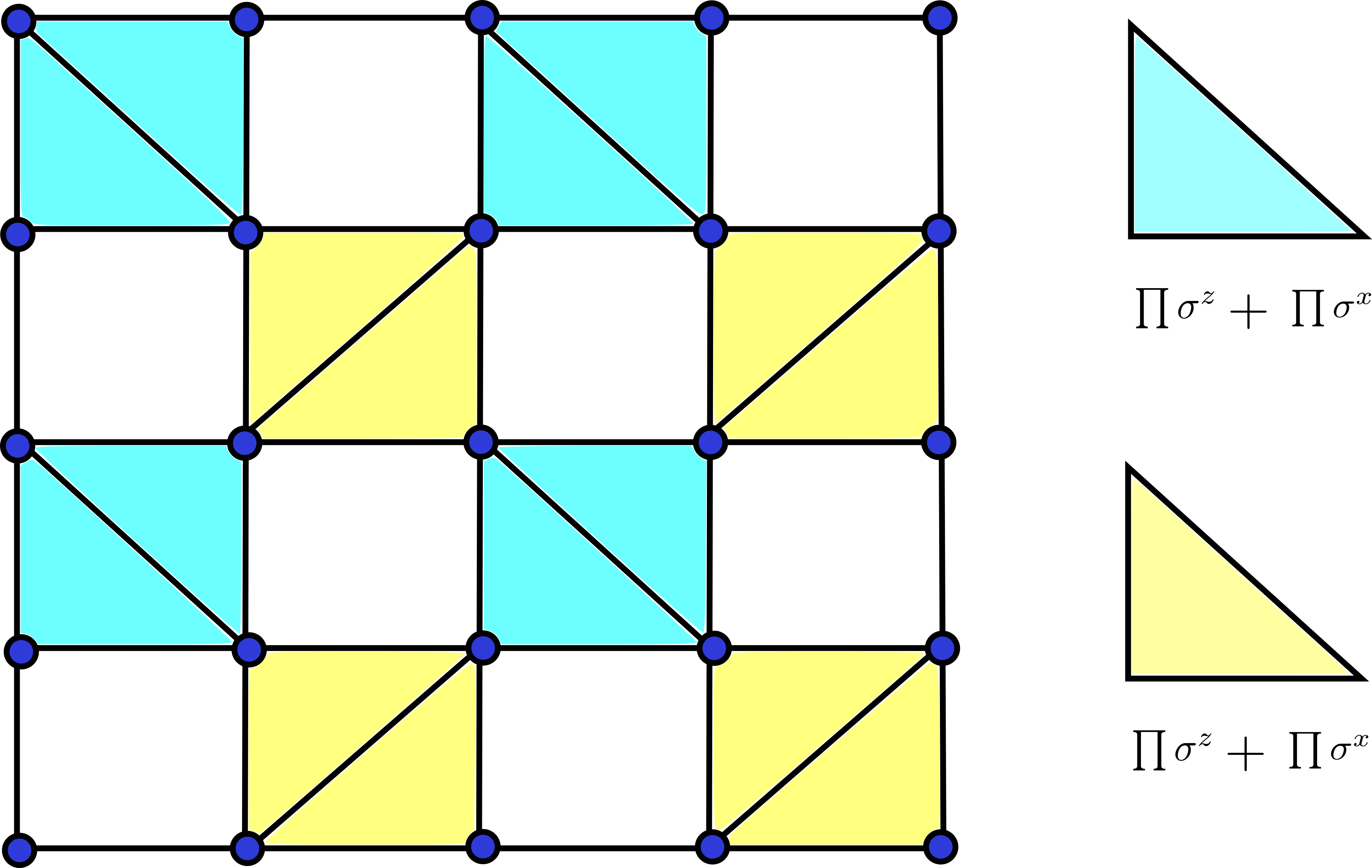}
    \caption{Shastry-Sutherland lattice with triangular plaquettes (shaded) supporting Hamiltonian terms, while square plaquettes are ``unoccupied''. The different colouring highlights neighbouring plaquettes that share a single site, illustrating the odd-overlap rule essential for the suppression of local symmetries. }
    \label{fig:SS}
\end{figure}

\begin{figure}
    \centering
    \includegraphics[width=\linewidth]{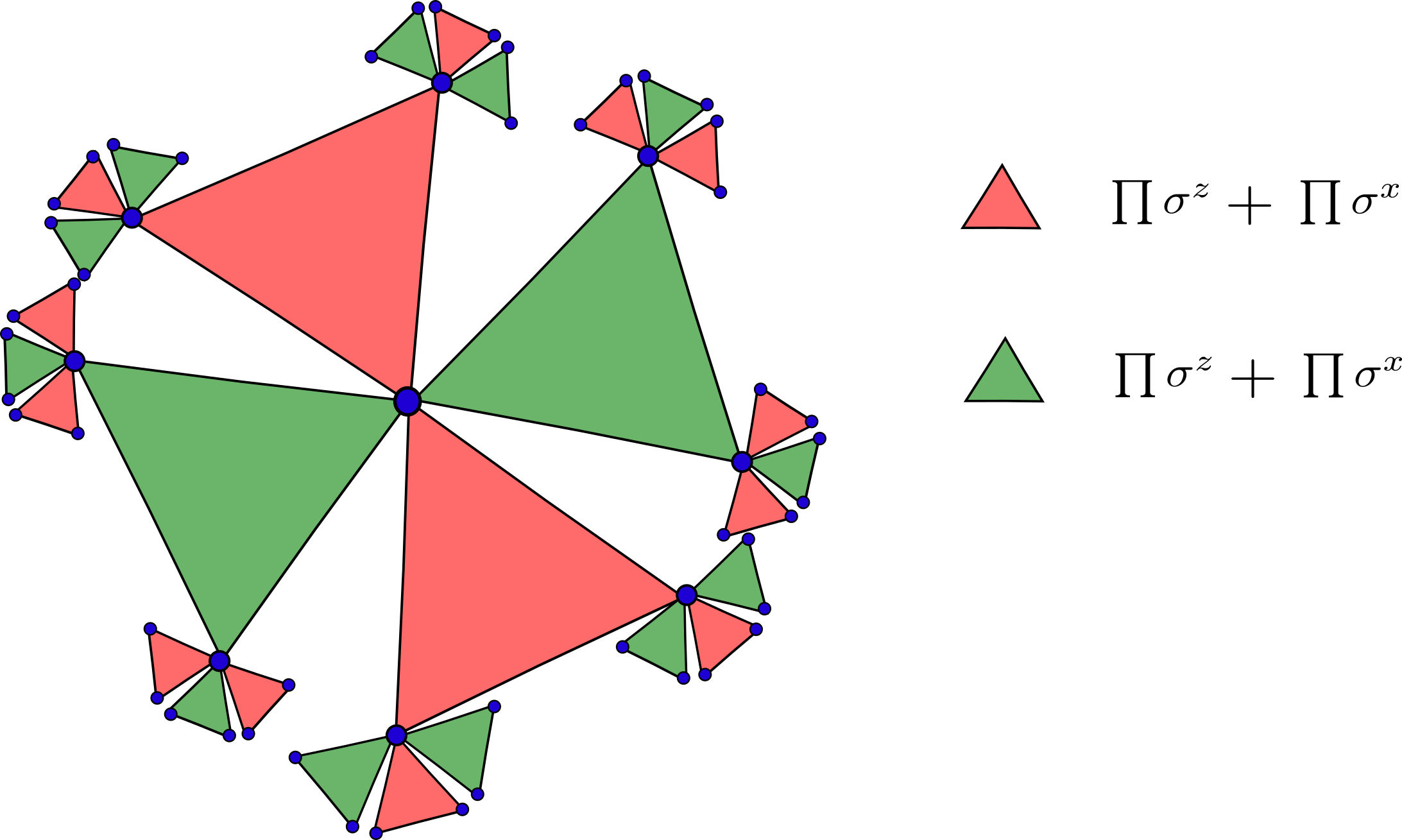} 
    \caption{A schematic illustration of a $z=4$ triangular cactus lattice. In this lattice, all plaquettes (red and green) have both $\prod_{i\in \triangle}\sigma_i^z$ and $\prod_{i\in \triangle}\sigma_i^x$ terms. Each plaquette in the model consists of an odd number of bonds, and any two plaquettes share only an odd number of sites. As a result, the model does not possess any local symmetries.}
    \label{fig:heptagon_kagome}
\end{figure}

Here we discuss models which avoid the \guillemotleft anticommuting\guillemotright~structure even though the lattices still possess corner-sharing properties similar to those discussed in the main text.
An instructive example of a lattice with a constrained local structure and no local symmetries can be found the Shastry-Sutherland lattice~\cite{SRIRAMSHASTRY19811069}. 
In this geometry, we define a Hamiltonian such that the corner-sharing square plaquettes remain unoccupied, while each pair of adjacent triangular plaquettes hosts two multi-spin coupling terms as shown in Fig.~\ref{fig:SS}. 
A key structural feature is that each occupied plaquette consists of an odd number of bonds (specifically three), and any two neighboring plaquettes of different types share an odd number of sites (typically one). 
This odd-overlap rule plays a crucial role in voiding local conserved quantities.

This principle can be extended to hierarchical or tree-like geometries. 
An illustrative example is the $z=4$  triangular cactus lattice shown in Fig.~\ref{fig:heptagon_kagome}. 
Here each site is connected to four triangular plaquettes in a decoration that avoids the formation of loops at small scales resulting in a locally tree-like structure. 
In this model, all the plaquettes are assigned two types of multi-spin coupling terms: $\prod_{i \in \triangle} \sigma^z_i$ and (2) $\prod_{i \in \triangle} \sigma^x_i$. As in the Shastry-Sutherland lattice, each plaquette contains an odd number of bonds, and neighbouring plaquettes of different types overlap on an odd number of sites. This geometric constraint enforces the absence of both local and extended conserved quantities.
Consequently, the cactus lattice offers a clean and geometrically controlled platform for constructing models with intrinsically constrained local dynamics without any local conserved quantities nor multi-Majorana form symmetries.


 \end{appendix}

\end{document}